\documentclass[11pt]{article}
\usepackage[utf8]{inputenc}
\usepackage[left=1in,right=1in]{geometry}
\usepackage{authblk}
\usepackage[pdfstartview=FitH,pdfpagemode=None]{hyperref}
\usepackage{amsmath}
\usepackage{amsfonts}
\usepackage{graphicx,graphbox}
\usepackage{subcaption}
\usepackage{braket}
\usepackage{comment}
\usepackage{cite}

\title{Reparameterization Dependence and Holographic Complexity of Black Holes}
\author[1,2]{Ayoub Mounim}
\author[1,2]{Wolfgang M\"uck}
\affil[1]{Dipartimento di Fisica ``Ettore Pancini", Universit\`a degli Studi di Napoli ``Federico II" \authorcr Via Cintia, 80126 Napoli, Italy}
\affil[2]{Istituto Nazionale di Fisica Nucleare, Sezione di Napoli \authorcr Via Cintia, 80126 Napoli, Italy}
\date{\today}

\begin{document}


\numberwithin{equation}{section}

\newcommand{\ie}{i.e.,\ }
\newcommand{\eg}{e.g.,\ }

\newcommand{\const}{\operatorname{const.}}

\newcommand{\sgn}{\operatorname{sgn}}

\newcommand{\rmd}{\,\mathrm{d}}

\newcommand{\Tr}{\operatorname{tr}}

\newcommand{\arctanh}{\operatorname{arctanh}}


\newcommand{\re}{\operatorname{Re}}
\newcommand{\im}{\operatorname{Im}}

\newcommand{\e}[1]{\operatorname{e}^{#1}}


\newcommand{\vev}[1]{\left\langle #1 \right\rangle}


\newcommand{\Order}{\mathcal{O}}

\newcommand{\unit}[1]{\operatorname{#1}}

\newcommand{\GR}[1]{\stackrel{\circ}{#1}{}\mspace{-5mu}}
\newcommand{\WB}[1]{\stackrel{\bullet}{#1}{}\mspace{-5mu}}

\maketitle
\begin{abstract}
We refine the calculation of holographic complexity of black holes in the complexity equals action approach by applying the recently introduced criterion that the action of any causal diamond in static vacuum regions must vanish identically. This criterion fixes empty anti-de Sitter (AdS) spacetime as the reference state with vanishing complexity and renders holographic complexity explicitly finite in all the cases we consider. The cases considered here include the Reissner-Nordström-AdS black hole, the rotating BTZ black hole, the Kerr-AdS black hole, and AdS-Vaidya spacetime. The criterion is equivalent to imposing that the corner contributions vanish. Contrary to earlier results, we find that the generalized Lloyd bound always holds in the Reissner-Nordström-AdS and BTZ cases. 
\end{abstract}
\section{Introduction}
\label{intro}

In recent years, the application of concepts and ideas of information theory to quantum field theory and gravity has been proven to be a fruitful line of research, especially in the context of the AdS/CFT correspondence\cite{Maldacena:1997re, Gubser:1998bc, Witten:1998qj, Ryu:2006ef, Ryu:2006bv, Casini:2011kv}. Holographic complexity is an example of this approach.  

In information theory, computational complexity is a measure of how difficult it is (or how many steps it takes) to approximately compute a desired target state starting from a given reference state and using a certain set of elementary operations \cite{Aaronson:2016vto}. By definition, the reference state is ``simple'' and has zero complexity. The definition of computational, or circuit, complexity depends on the system under consideration, the set of elementary operations, the reference state and a parameter $\epsilon$ that specifies the tolerance with which the target state is reached. Typically, the computational complexity diverges when $\epsilon \to 0$. A geometric approach to complexity, which can be applied to quantum field theory, was developed in \cite{Nielsen1133}, defining complexity in terms of a weight function evaluated on a trajectory connecting the target and the reference state in some space of unitary operators. Several proposals for the weight function have been investigated in \cite{Jefferson:2017sdb, Chapman:2017rqy, Sinamuli:2019utz, Yang:2017nfn, Khan:2018rzm, Doroudiani:2019llj}. 

A different notion of complexity is operator, or Krylov, complexity \cite{Parker:2018yvk}. Operator complexity is a measure of how much a given operator spreads out in the space of operators under Heisenberg evolution. It is a function of time and depends on the system under consideration, the choice of an inner product in the space of operators and the initial operator. By definition, the initial operator is simple, \ie operator complexity vanishes initially. In contrast to computational complexity, operator complexity is intrinsically finite. 

Given the variety of complexity measures that one can define in quantum and information theory, it is no surprise that several gravitational observables in asymptotically Anti-de Sitter (AdS) spacetime have been proposed as holographic duals of complexity. These proposals fall into two classes.\footnote{A new infinite family of observables that are viable as gravitational duals of complexity has been defined in \cite{Belin:2021bga}.}
The first proposal is known as the "complexity equals volume" ($C=V$) approach \cite{Susskind:2014rva, Alishahiha:2015rta}, which derives from the observation that the interior of a black hole continues to grow linearly for an exponentially long time after the black hole has formed. The second proposal is the "complexity equals action" ($C=A$) \cite{Brown:2015bva, Brown:2015lvg} approach, in which complexity is identified with the action evaluated in a bulk region called the Wheeler-de Witt (WdW) patch,\footnote{In the rest of the paper, we will work with the reduced action, $I = 16\pi G S$.}
\begin{equation}
\label{intro:complexity}
\mathcal{C} = \frac{S_{\mathrm{WdW}}}{\pi\hbar}~.
\end{equation}
The WdW patch is defined as the region bounded by the null surfaces anchored at certain times on the spacetime boundary (left and right boundaries in the case of two-sided black holes) and, possibly, the black hole singularity.

In a recent paper \cite{Mounim:2021bba}, we have proposed a refinement of the $C=A$ approach introducing the criterion that the action in any causal diamond in static vacuum regions should vanish. 
With this criterion, the complexity of the state dual to pure AdS spacetime vanishes by construction, which identifies this state as the reference state. Moreover, because the asymptotic region does not contribute when this criterion is applied, the holographic complexity of AdS-Schwarzschild black holes turns out to be finite \cite{Mounim:2021bba}.   
The freedom to introduce this criterion derives from the fact that the minimal action terms on the null boundaries of the WdW patch, which are required by the variational principle, are not reparameterization invariant. To be more precise, if $\Phi=0$ specifies a null boundary, \ie the vector $\partial_\mu \Phi$ is null, then the tangent vector along the null direction is given by
\begin{equation}
\label{intro:k}
	k^\mu =\frac{\partial x^\mu}{\partial\lambda} = \e{\sigma} \partial^\mu \Phi~,
\end{equation}
where $\sigma$ can be an arbitrary function of the intrinsic coordinates on the null hypersurface, amongst which $\lambda$ parameterizes the null direction. It is known \cite{Lehner:2016vdi} that the action depends on the choice of $\sigma$. Reparameterization dependence is typically regarded as unphysical. Therefore, the usual approach to avoiding this ambiguity is to add a counter term to the null boundary action, which does not interfere with the variational principle \cite{Lehner:2016vdi} and renders the full action reparameterization invariant. Although adding such a term is in line with the principles of holographic renormalization \cite{Emparan:1999pm, deHaro:2000vlm, Bianchi:2001kw, Martelli:2002sp, Skenderis:2002wp} and has the virtue of facilitating the calculation allowing simple choices of $\sigma$, it does not make the action finite, nor does adding any other covariant boundary term.

Instead, the approach taken in \cite{Mounim:2021bba} aims to identify a privileged choice of parameterization by imposing the criterion of  vanishing action on any static vacuum causal diamond. This criterion defines the state dual to empty AdS space as the reference state, because it has zero complexity by construction. In addition, the complexity of AdS-Schwarzschild spacetime was found to be finite. As a function of time $\tau$, it remains constant (equal to the complexity of formation) from $\tau=0$ up to a certain critical time $\tau_c$ and grows linearly thereafter, with a growth rate saturating the Lloyd bound \cite{Lloyd:2000, Brown:2015lvg}. We remark that looking for a privileged class of parameterizations is justified, because the non-invariant null boundary terms carry a physical meaning as the heat flux through the boundary \cite{Chakraborty:2019doh}, so that different parameterizations may describe physically different situations. 

In the present article, we follow up on our initial proposal \cite{Mounim:2021bba} and reconsider holographic complexity in the $C=A$ approach in the cases of Reissner-Nordström-AdS (RN-AdS) black holes, the rotating BTZ black hole, the Kerr-AdS black holes, and AdS-Vaidya spacetime.
All of these black holes have been considered before, for example, in  \cite{Brown:2015bva, Brown:2015lvg, Lehner:2016vdi, Carmi:2017jqz, Akhavan:2018wla, Akhavan:2019zax, Swingle_2018, Cano_2018, Auzzi_2018, Yaraie_2018, Pan_2017, Goto_2019, Balushi:2020wjt, Hashemi:2019aop, Alishahiha:2019cib, Alishahiha:2018tep, Alishahiha:2018lfv, Omidi:2020oit, Bernamonti:2021jyu}, so that our work is not entirely new. What is new, though, is the choice of parameterization of the null boundaries and the fact that the counter term is deliberately omitted. In particular, we will demonstrate that the complexity is finite in all of the cases we consider. We will also investigate whether or not the complexity growth rate satisfies Lloyd's bound or a suitable generalization thereof \cite{Cai:2016xho}. In the cases of the RN-AdS and rotating BTZ black holes, the answer will be affirmative for the generalized bound \cite{Cai:2016xho}. These results contradict the findings of \cite{Carmi:2017jqz, Auzzi_2018, Bernamonti:2021jyu}, which also shows that our approach is an improvement of the $C=A$ proposal. In the case of Kerr-AdS, we are not able to give a definite answer, but we can establish that the limiting value is approached from below at late times.\footnote{For AdS-Vaidya, the validity of the bound follows from AdS-Schwarzschild.} 

The rest of the paper is organized as follows. For the sake of brevity, we avoid repeating the details regarding the action in the WdW patch and refer readers to section~2 of \cite{Mounim:2021bba}, also for what concerns our notation. 
In section~\ref{rna}, we compute the complexity of the charged Reissner-Nordström-AdS black hole. As examples for rotating black holes, the rotating BTZ solution and Kerr-AdS spacetime are considered in sections~\ref{btz} and~\ref{kads}, respectively. In section~\ref{vda}, we study the complexity of Vaidya spacetime, which describes the formation of a spherically symmetric black hole by gravitational collapse of a null fluid. Finally, we conclude in section~\ref{conc}.

\section{Reissner-Nordström-AdS black hole}
\label{rna}

\subsection{Setup}

In this section, we will compute the complexity of an (electrically charged) RN-AdS spacetime with respect to empty AdS. Because of the presence of the electric field, there is no vacuum region in RN-AdS. We will first compute the action in a generic causal diamond, and then use the criterion introduced in \cite{Mounim:2021bba} to pick a specific parameterization.

RN-AdS spacetime is a solution of the Maxwell-Einstein theory defined by the action 
\begin{equation}
    \label{rna:action}
    I = \int\rmd x^{n+2}\sqrt{-g}\left(R - 2\Lambda - F_{\mu\nu}F^{\mu\nu}\right)~.
\end{equation}
The solution is given by an electric potential\footnote{$r_+$ denotes the outer horizon radius, see below.}
\begin{equation}
    \label{rna:pot}
    A_\mu\rmd x^\mu = q\sqrt{\frac{n}{2(n-1)}}\left(\frac{1}{r_+^{n-1}}-\frac{1}{r^{n-1}}\right)\rmd t~,
\end{equation}
and the metric
\begin{equation}
\label{rna:metric}
\rmd s^2 = -f(r) \rmd t^2 +f(r)^{-1} \rmd r^2 + r^2 \rmd \Omega_n^2~,
\end{equation}
with the blackening function $f(r)$ defined by
\begin{equation}
\label{rna:f}
 f(r) = 1 + \frac{r^2}{L^2} - \frac{\omega^{n-1}}{r^{n-1}} + \frac{q^2}{r^{2(n-1)}}~.
\end{equation}
The parameter $\omega$ is related to the total mass\footnote{$\Omega_n$ denotes the volume of a unit $n$-sphere.} 
\begin{equation}
\label{rna:M}
	M = \frac{n\Omega_n}{16\pi G} \omega^{n-1}~,
\end{equation} 
while the parameter $q$ determines the electric charge of the black hole
\begin{equation}
    \label{rna:charge}
    Q = q\sqrt{2n(n-1)}\frac{\Omega_n}{8\pi G}~.
\end{equation}
The horizon radii are defined by the zeros of $f(r)$. Let us, for the moment, consider the non-extremal case, in which there are two horizons at $r = r_+$ and $r = r_- < r_+$). The relevant thermodynamic variables are associated with the outer horizon, $r_+$. The chemical potential, temperature, and entropy are given by
\begin{align}
	\mu &= \sqrt{\frac{n}{2(n-1)}}\frac{q}{r_+^{n-1}}~,\\
	T &= \frac{1}{4\pi}\left(\frac{2r_+}{L^2} + \frac{(n-1)\omega^{n-1}}{r_+^n} - \frac{2(n-1)q^2}{r_+^{2n-1}}\right)~,\\
	S &= \frac{\Omega_nr_+^n}{4G}~,
\end{align}
respectively.

\begin{figure}[t]
	\begin{center}
		\includegraphics[width=.20\textwidth]{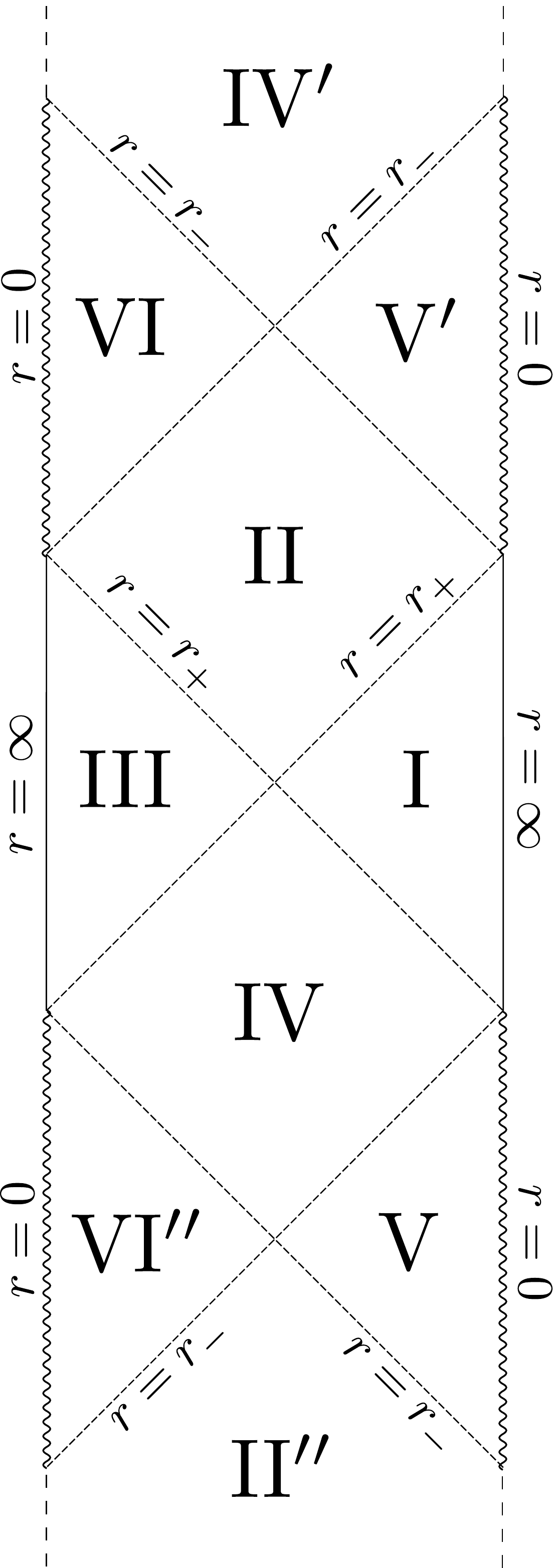}%
		\caption{Penrose diagram of the RN-AdS black hole. The diagram repeats itself periodically above and below.\label{fig:RN.Pd}}
	\end{center}
\end{figure}

The Penrose diagram of part of the extended (non-extremal) RN-AdS spacetime is shown in figure~\ref{fig:RN.Pd}. Each of the numbered regions is covered by a set of coordinates $(t,r)$ with metric \eqref{rna:metric}, with $r>r_+$ in regions $\mathrm{I}$ and $\mathrm{III}$, $r_-<r<r_+$ in $\mathrm{II}$ and $\mathrm{IV}$, and $0<r<r_-$ in regions $\mathrm{V}$ and $\mathrm{VI}$. The curvature singularities are situated at $r=0$.
Similarly to the AdS-Schwarzschild black hole case, it will be useful to work with Eddington-Finkelstein coordinates. The tortoise coordinate can be defined by
\begin{equation}
\label{rna:tortoise}
	r^\ast(r) = \int\limits_{R}^r \frac{\rmd r}{f(r)}~,
\end{equation} 
where $R$ is identified with the cut-off radius in the asymptotic region, which will be sent to $\infty$ at the end. 
With ingoing Eddington-Finkelstein coordinates, $v = t+ r^\ast$, the metric is 
\begin{equation}
\label{rna:metricVR}
\rmd s^2 = -f(r)\rmd v^2 + 2\rmd v\rmd r + r^2\rmd\Omega_n^2~.
\end{equation}
The ingoing Eddington-Finkelstein coordinate patches extend over three numbered regions, $\mathrm{I}\cup\mathrm{II}\cup\mathrm{VI}$, $\mathrm{III} \cup \mathrm{IV}\cup\mathrm{V}$, or any of their periodic repetitions.

Likewise, with outgoing coordinates, $u = t -r^\ast$, the metric is 
\begin{equation}
\label{rna:metricUR}
\rmd s^2 = -f(r)\rmd u^2 - 2\rmd u\rmd r + r^2\rmd\Omega_n^2~.
\end{equation}
The outgoing Eddington-Finkelstein coordinates cover the patches $\mathrm{I} \cup\mathrm{IV}\cup\mathrm{VI''}$, $\mathrm{II} \cup \mathrm{III}\cup\mathrm{V'}$, or any of their periodic repetitions.

\subsection{Action in a causal diamond}
\begin{figure}[t]
	\begin{center}
		\includegraphics[width=.4\textwidth]{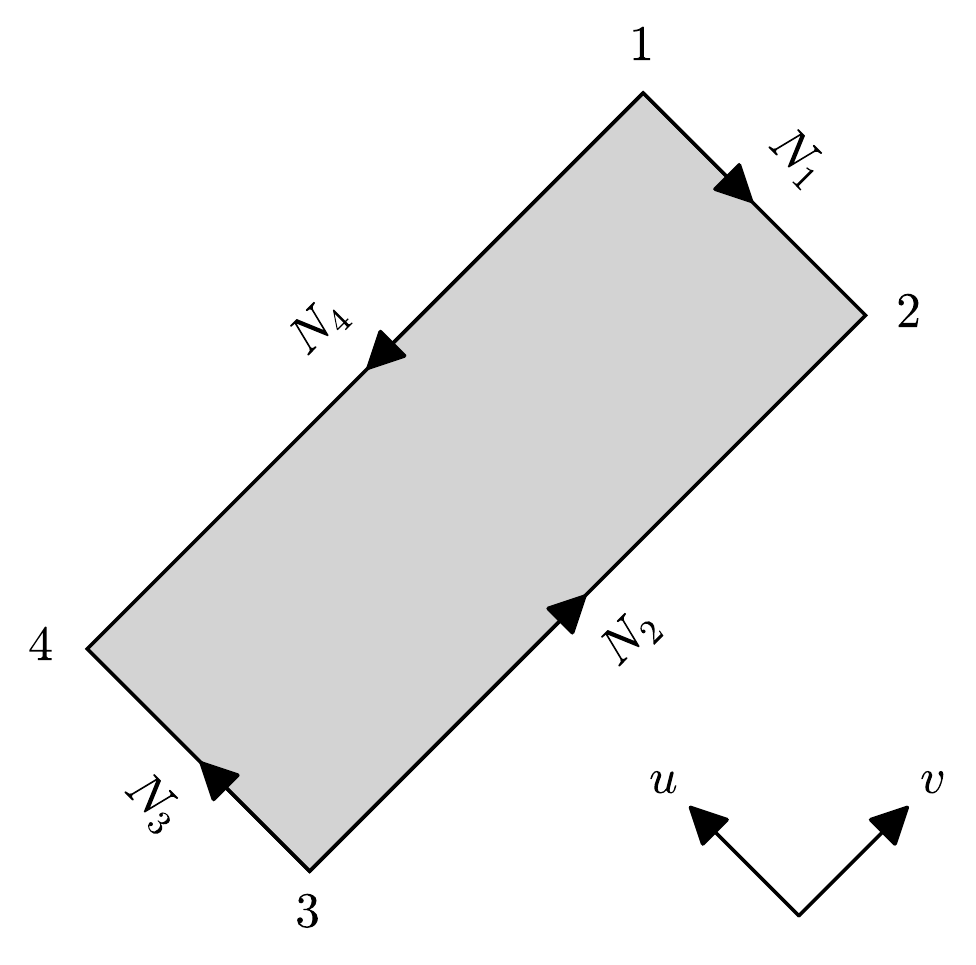}%
		\caption{Setup of the causal diamond computation.
		The labels of the null boundaries and the corners used in the text are shown.
		\label{fig:caus.diamond}}
	\end{center}
\end{figure}
We focus on a generic causal diamond embedded in RN-AdS spacetime. A causal diamond is bounded by four null surfaces, which we label $N_1$, \ldots, $N_4$, counting them clockwise starting from the north east. The four intersection points are counted clockwise starting from the north.  Obviously, their coordinates satisfy $v_1 = v_2$, $u_2=u_3$, $v_3=v_4$ and $u_4 = u_1$. 
The setup is illustrated in figure~\ref{fig:caus.diamond}.
In what follows, we work in outgoing Eddington-Finkelstein coordinates \eqref{rna:metricUR}. 

The four scalar functions defining the null surfaces are given by
\begin{subequations}
\begin{align}
	\label{rna:phi1}
		\Phi_1(u,r) &= u+ 2r^\ast(r) - v_1~,\\
	\label{rna:phi2}
		\Phi_2(u,r) &= u_2 -u~,\\
	\label{rna:phi3}
		\Phi_3(u,r) &= v_3 - u- 2r^\ast(r)~,\\
	\label{rna:phi4}
		\Phi_4(u,r) &= u - u_4~.
\end{align}
\end{subequations}
From these, we obtain the following expressions for the null tangent vectors,
\begin{subequations}
\begin{align}
	\label{rna:k1}
		k_1^\alpha &= \e{\sigma_1} \left( -\frac{2}{f}, 1 , \vec{0} \right)~,\\
	\label{rna:k2}
		k_2^\alpha &= \e{\sigma_2} \left( 0, 1 , \vec{0} \right)~,\\
	\label{rna:k3}
		k_3^\alpha &= \e{\sigma_3} \left( \frac{2}{f}, -1 , \vec{0} \right)~,\\
	\label{rna:k4}
		k_4^\alpha &= \e{\sigma_4} \left( 0, -1 , \vec{0} \right)~.
\end{align}
\end{subequations}
The four functions $\sigma_1$, \ldots, $\sigma_4$ implement the parameterization dependence. 

Let us start with the surface terms. The orientation of the $\lambda$-integrals can be read off from \eqref{rna:k1}--\eqref{rna:k4}, because $k^\alpha = \partial_\lambda x^\alpha$. From \eqref{rna:k1}--\eqref{rna:k4} it can also be shown that the non affinity parameter on each null boundary is $\kappa_i=\partial_\lambda \sigma_i$. 
Computing the contribution of the four boundaries, we find, after an integration by parts
\begin{align}
\label{rna:I.N}
	\frac{I_{N}}{2\Omega_n} &= 
	-n\int\limits_{r_1}^{r_2}\rmd r\,r^{n-1}\sigma_1
	-n\int\limits_{r_3}^{r_2}\rmd r\,r^{n-1}\sigma_2 
	-n\int\limits_{r_3}^{r_4}\rmd r\,r^{n-1}\sigma_3 
	-n\int\limits_{r_1}^{r_4}\rmd r\,r^{n-1}\sigma_4 \\
\notag
	&\quad 
	- r_1^n\left[\sigma_1(r_1)+\sigma_4(r_1)\right] + r_2^n\left[\sigma_1(r_2)+\sigma_2(r_2)\right] 
	- r_3^n\left[\sigma_2(r_3)+\sigma_3(r_3)\right] + r_4^n\left[\sigma_3(r_4)+\sigma_4(r_4)\right]~.
\end{align}

We now consider the bulk contribution. From \eqref{rna:pot} we have
\begin{equation}
    \label{rna:maxwell}
    F_{\mu\nu}F^{\mu\nu} = -\frac{n(n-1)q^2}{r^{2n}}~,
\end{equation}
while from Einstein's equation
\begin{equation}
    \label{rna:einstein}
    R = \frac{2(n+2)}{n}\Lambda + \frac{n-2}{n}F_{\mu\nu}F^{\mu\nu}~.
\end{equation}
Thus, the on-shell action \eqref{rna:action} reads
\begin{align}
    \notag
    I_B &= \Omega_n\int\rmd r\rmd u\,r^n\left[-\frac{2(n+1)}{L^2} + \frac{2(n-1)q^2}{r^{2n}}\right]\\
    \notag
    &= 2\Omega_n\int\limits_{u_2}^{u_4}\rmd u\left[\frac{1}{L^2}\left(\rho_3^{n+1}-\rho_1^{n+1}\right)+q^2
    \left(\rho_3^{1-n}-\rho_1^{1-n}\right)\right]\\
\label{rna:IB-2} 
    &= 2\Omega_n\int\limits_{u_2}^{u_4}\rmd u\,\left[\rho_1^n\frac{\rmd\ln |f|}{\rmd u} - \rho_3^n\frac{\rmd \ln |f|}{\rmd u} + n q^2
    \left(\rho_3^{1-n}-\rho_1^{1-n}\right)\right]~.
\end{align}
The functions $\rho_1(u)$ and $\rho_3(u)$ are defined implicitly by $\Phi_1(u,\rho_1)=0$ and $\Phi_3(u,\rho_3)=0$, respectively, and in the last line we have used the identity
\begin{equation}
    \label{rna:id}
    \frac{\rho}{L^2} = -\frac{\rmd}{\rmd u} \ln |f|- \frac{(n-1)\omega^{n-1}}{2\rho^n} + \frac{q^2(n-1)}{\rho^{2n-1}}~.
\end{equation}
After integrating by parts and changing the integration variable, \eqref{rna:IB-2} becomes
\begin{align}
\label{rna:IB-3} 
	\frac{I_{B}}{2\Omega_n} 
		&= n\int\limits_{r_1}^{r_2} \rmd r\, r^{n-1} \ln |f| 
		+ n\int\limits_{r_3}^{r_4} \rmd r\, r^{n-1} \ln |f| 
		-2n q^2\int\limits_{r_3}^{r_4}\rmd r\frac{r^{1-n}}{f}
		-2n q^2\int\limits_{r_1}^{r_2}\rmd r\frac{r^{1-n}}{f}\\
\notag &\quad
		+ r_1^n \ln |f(r_1)| - r_2^n \ln |f(r_2)|
		+ r_3^n \ln |f(r_3)| - r_4^n \ln |f(r_4)|~.
\end{align}
This is identical to 
\begin{align}
\label{rna:IB}
	\frac{I_B}{2\Omega_n} 
		&= n a_u\int\limits_{r_1}^{r_2}\rmd r\,r^{n-1}\ln |f| 
		 + n a_u\int\limits_{r_3}^{r_4}\rmd r\,r^{n-1}\ln |f| \\
	\notag &\quad
		+ n a_v\int\limits_{r_3}^{r_2}\rmd r\,r^{n-1}\ln |f| 
		+ n a_v\int\limits_{r_1}^{r_4}\rmd r\,r^{n-1}\ln |f| \\
	\notag &\quad 
	-2nq^2\int\limits_{r_3}^{r_4}\rmd r\frac{r^{1-n}}{f}
	-2nq^2\int\limits_{r_1}^{r_2}\rmd r\frac{r^{1-n}}{f}\\
	\notag &\quad
	+ r_1^n\ln |f(r_1)| - r_2^n\ln |f(r_2)|+r_3^n\ln |f(r_3)|-r_4^n\ln |f(r_4)|~,
\end{align}
where $a_u$ and $a_v$ are two real constants that are constrained by $a_u + a_v = 1$.

The corner terms contribute
\begin{align}
\label{rna:IC}
	\frac{I_C}{2\Omega_n} 
		&= r_1^n \left[\sigma_1(r_1)+\sigma_4(r_1)-\ln |f(r_1)|\right] 
		- r_2^n \left[\sigma_1(r_2)+\sigma_2(r_2)-\ln |f(r_2)|\right] \\ 
\notag &\quad 
		+ r_3^n \left[\sigma_2(r_3)+\sigma_3(r_3)-\ln |f(r_3)|\right] 
		- r_4^n \left[\sigma_3(r_4)+\sigma_4(r_4)-\ln |f(r_4)|\right]~.
\end{align}
In the above equations, the manipulations we have done are such that the terms arising from the integration by parts of the surface and bulk  contributions precisely cancel the corner terms. We now choose a particular parameterization of the null boundary hypersurfaces by specifying the parameterization functions $\sigma_i(r)$. The choice is driven by analogy to the Schwarzschild-AdS case discussed in \cite{Mounim:2021bba}, where, in order to measure the complexity of the black hole with respect to empty AdS space, the parameterization functions turned out to be proportional to the logarithm of the blackening factor of the black hole. Concretely, 
\begin{equation}
\label{rna:choice}
\begin{aligned}
\text{on $N_1$:}\quad \sigma_1(\lambda) &= a_u \ln |f(r(\lambda))|~, \qquad 
\text{on $N_3$:}\quad \sigma_3(\lambda)  = a_u \ln |f(r(\lambda))|~,\\
\text{on $N_2$:}\quad \sigma_2(\lambda) &= a_v \ln |f(r(\lambda))|~, \qquad 
\text{on $N_4$:}\quad \sigma_4(\lambda)  = a_v \ln |f(r(\lambda))|~.
\end{aligned}
\end{equation}
With this choice, we note that also the corner term contribution \eqref{rna:IC} vanishes identically. Adding \eqref{rna:IB} and \eqref{rna:I.N} yields the total action of the causal diamond
\begin{align}
\label{rna:CDfix}
	I &= 4nq^2\Omega_n\left[\int\limits_{r_4}^{r_3}\rmd r\frac{r^{1-n}}{f} + \int\limits_{r_2}^{r_1}\rmd r\frac{r^{1-n}}{f}\right]~.
\end{align}
Clearly, when the electric charge of the black hole is set to zero while keeping fixed the positions $r_i$ of the four corners of the causal diamond, the action vanishes, as required by our criterion. However, this does not imply that the $q\to0$ limit results in a vanishing complexity, because $\tau$ should be held fixed in this limit, not the $r$-variables of the corners.

\subsection{Complexity=Action}

\begin{figure}[t]
	\begin{center}
		\includegraphics[width=.3\textwidth]{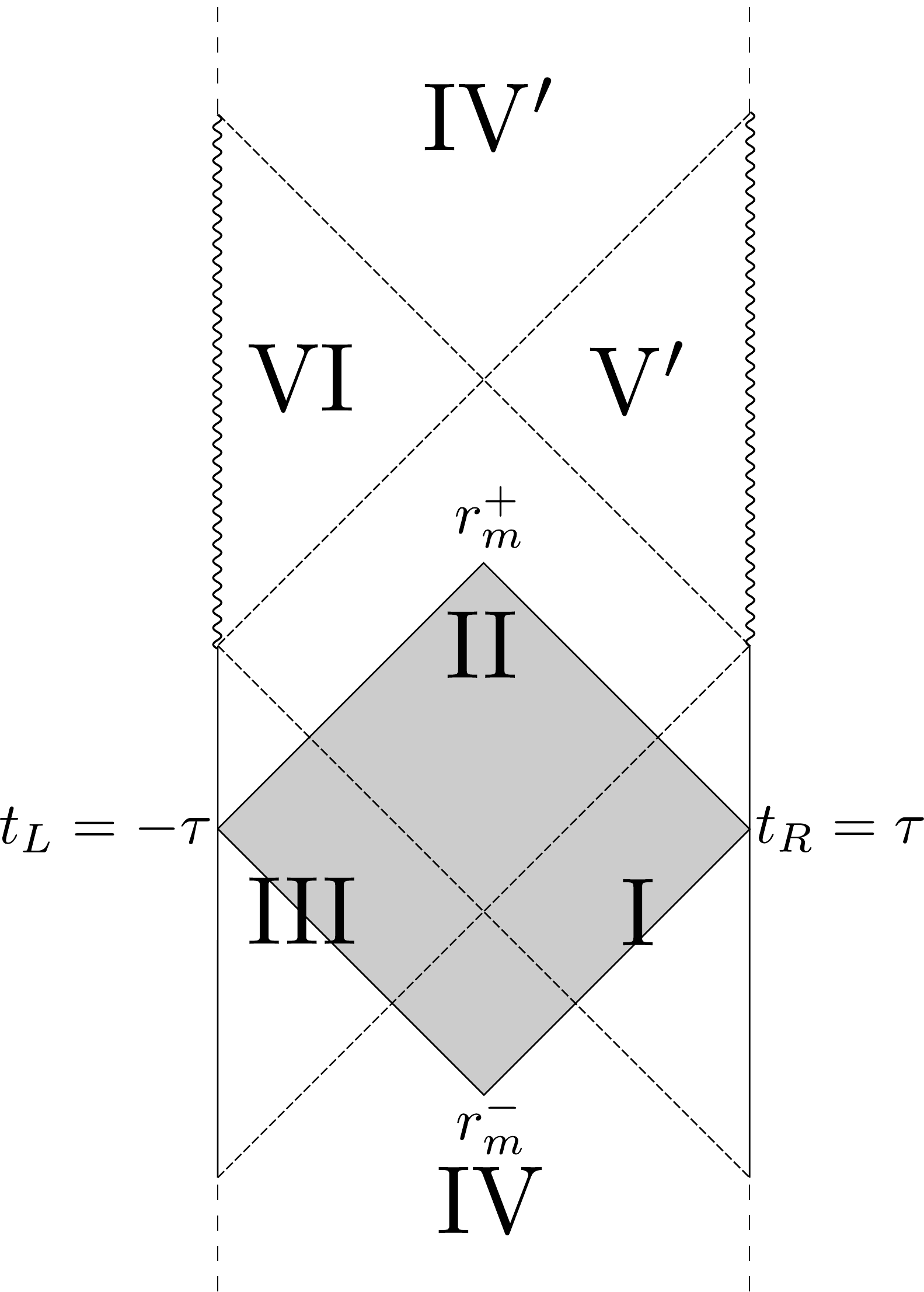}%
		\caption{WdW patch of the RN-AdS black hole.\label{fig:RN.WdW}}
	\end{center}
\end{figure}

The WdW patch is bounded by the null surfaces intersecting the left and right boundaries at the cut-off radius $R$ at times $t_L$ and $t_R$, respectively. This is illustrated in figure~\ref{fig:RN.WdW}. To simplify, we can use time translation invariance to set $t_R=-t_L=\tau$. 
Contrary to the AdS-Schwarzschild case, the WdW patch does not reach the singularity.\footnote{The extremal RN-AdS black hole, which has a different causal structure, is an exception. We shall comment on it at the end of this section.} Therefore, it is always a causal diamond. The future and past vertices of this diamond are located in the regions $\mathrm{II}$ and $\mathrm{IV}$, respectively, at the radii $r=r_m^+$ and $r=r_m^-$ determined by 
\begin{equation}
	\label{rna:rm}
	r^\ast(r_m^\pm) = \pm\tau~, \qquad (r_- < r_m^\pm < r_+)~. 
\end{equation}
This implies $r_m^- = r_m^+$  for $\tau=0$, and $r_m^\pm\to r_\mp$ for $\tau\to \infty$. We restrict our attention to $\tau>0$. 

The result \eqref{rna:CDfix} is straightforwardly translated into complexity using \eqref{intro:complexity}. Substituting the radial positions of the four corners of the WdW patch, one finds 
\begin{equation}
    \label{rna:I.wdw}
    \mathcal{C} 
    =-\frac{n\Omega_nq^2}{4\pi^2G\hbar}\left[\int\limits_{r_m^+}^{r_m^-}\rmd r\,\frac{r^{1-n}}{f} + 2\int\limits_{r_m^-}^{R}\rmd r\,\frac{r^{1-n}}{f}\right]~.
\end{equation}
The integrand in the second term of \eqref{rna:I.wdw} has an integrable singularity at the outer horizon, $r = r_+$. Moreover, the second term is also integrable in the $R\to\infty$ limit, so that the cut-off $R$ can be safely removed. In other words, the complexity is finite in our approach.

The time $\tau$ appears in the complexity only implicitly through the integration limits $r_m^\pm$. 
Using \eqref{rna:rm} and \eqref{rna:tortoise}, the complexity growth rate is found to be  
\begin{equation}
    \label{rna:I.der}
    \frac{\rmd\mathcal{C}}{\rmd\tau} = \frac{n\Omega_nq^2}{4\pi^2G\hbar}\left[\left(r_m^+\right)^{1-n} - \left(r_m^-\right)^{1-n}\right]. 
\end{equation}
This vanishes at $\tau = 0$, is always positive for $\tau > 0$, and approaches, at late times, the value known from the literature \cite{Brown:2015lvg, Cai:2016xho, Lehner:2016vdi, Jiang:2019qea},
\begin{equation}
    \label{rna:I.derlimit}
    \lim_{\tau\rightarrow\infty}\frac{\rmd\mathcal{C}}{\rmd\tau} = \frac{n\Omega_nq^2}{4\pi^2G\hbar}\left[\left(r_-\right)^{1-n} - \left(r_+\right)^{1-n}\right]~.
\end{equation}

In contrast to earlier results \cite{Carmi:2017jqz}, this limiting value is approached from below. To see this, calculate 
the second time derivative of the complexity,
\begin{equation}
	\label{rna:I.der2}
	\frac{\rmd^2\mathcal{C}}{\rmd\tau^2} = -\frac{n(n-1)\Omega_nq^2}{4\pi^2G\hbar}\left[\frac{f(r_m^-)}{(r_m^-)^{n}} + \frac{f(r_m^+)}{(r_m^+)^n}\right]~.
\end{equation} 
Because $f(r_m^-)$ and $f(r_m^+)$ are both negative for all times, \eqref{rna:I.der2} is always positive, and the complexity growth rate can only approach \eqref{rna:I.derlimit} from below.

This has an implication for the validity of the (generalized) Lloyd bound.
It has been observed \cite{Brown:2015lvg} that the late-time value \eqref{rna:I.derlimit} violates the conjectured Lloyd bound\footnote{We have included a factor of two in order to adjust for our differing time convention relative to \cite{Brown:2015lvg}.}
\begin{equation}
\label{rna:Lloyd.conj}
	\frac{\rmd\mathcal{C}}{\rmd\tau} \leq \frac{4}{\pi \hbar} \left[ (M-\mu Q) - (M-\mu Q)_{gs} \right]~,
\end{equation}
where "$gs$" stands for the ground state. In \cite{Cai:2016xho} it was proposed to consider the limiting value \eqref{rna:I.derlimit} as the Lloyd bound and reinterpret it as 
\begin{equation}
\label{rna:Lloyd.2}
	\frac{\rmd\mathcal{C}}{\rmd\tau} \leq \frac{2}{\pi \hbar} \left[ (M-\mu_+ Q) - (M-\mu_- Q) \right]~,
\end{equation}
where $\mu_\pm$ are the chemical potentials associated with the outer and inner horizons. If one adopts this viewpoint, our result implies  that the Lloyd bound is never violated, in contrast to the findings of \cite{Carmi:2017jqz}, where it was violated at intermediate times. This improvement is a direct result of our parameterization of the null boundaries.

A short comment is in order for the extremal case. The causal structure of the extremal black hole is different from the non-extremal cases. In particular, the extremal black hole is one-sided, \ie the WdW patch is anchored only on one asymptotic boundary. This, together with time translation symmetry, implies that the action of the WdW patch is time-independent. As a consequence, extremal black holes do not complexify \cite{Brown:2015lvg}. This agrees with taking the extremal limit of our results above. In particular, in the extremal limit the interval in which $f(r)<0$ shrinks to a point, so that all of $r_+$, $r_-$, $r_m^+$ and $r_m^-$ approach a common value, which implies that \eqref{rna:I.der}, \eqref{rna:I.derlimit} and \eqref{rna:I.der2} all vanish.

\section{Rotating BTZ black hole}
\label{btz}

\subsection{Setup}

Black holes without electric or magnetic charges, but with angular momentum, are vacuum solutions, but they are not static. Therefore, they provide another interesting testing case for our approach. As a first example, we shall consider the rotating BTZ black hole. The BTZ black hole is a solution of Einstein gravity with a negative cosmological constant in $(2+1)$-dimensions. The metric defining the geometry is\cite{Banados:1992wn}
\begin{align}
\label{btz:metric}
\rmd s^2 &= -f(r)\rmd t^2 + \frac{\rmd r^2}{f(r)} + r^2\left(\rmd\phi - \frac{j}{2r^2}\rmd t\right)^2~,
\end{align}
where
\begin{equation}
\label{btz:defs}
f(r) = \frac{r^2}{L^2} + \frac{j^2}{4r^2} - m~,
\end{equation}
is the blackening function of the black hole.
The physical mass and angular momentum of the black hole are $M = \frac{m}{8G}$ and $J = \frac{j}{8G}$. The causal structure of the BTZ black hole is similar to that of the RN-AdS black hole, with a singularity at $r = 0$ and two horizons defined by $f(r_\pm) = 0$. The horizon radii are
\begin{equation}
	r^2_\pm = \left(1\pm \sqrt{1-\frac{j^2}{m^2L^2}}\right)\frac{mL^2}{2}~.
\end{equation}
In order to have two distinct positive solutions, the angular momentum and the mass of the black hole must satisfy the relation $J<ML$. The special case $J=ML$ is the extremal case, in which the black hole has a single horizon, while, if $J>ML$, the space time defined by the metric \eqref{btz:metric} has a naked singularity. We will always assume $J<ML$ in the rest of the section.

To study the black hole, we will use Eddington-Finkelstein coordinates. We start by defining the tortoise-like coordinates
\begin{equation}
\label{btz:tortoise}
\begin{aligned}
r^\ast(r) &= \int\limits_R^r\frac{\rmd r}{f} = \frac{1}{2\sqrt{m^2-\frac{j^2}{L^2}}}\left(r_+\ln\frac{|r-r_+|}{r+r_+} - r_-\ln\frac{|r-r_-|}{r+r_-}\right)~,\\
r^\sharp(r) &= \int\limits_R^r\rmd r\frac{j}{2r^2f} = \frac{j}{4\sqrt{m^2-\frac{j^2}{L^2}}}\left(\frac{1}{r_+}\ln\frac{|r-r_+|}{r+r_+} - \frac{1}{r_-}\ln\frac{|r-r_-|}{r+r_-}\right)~.
\end{aligned}
\end{equation}
Then, the ingoing null coordinates are
\begin{equation}
\label{btz:ingoing}
\begin{aligned}
v &= t + r^\ast(r),\\
\psi &= \phi + r^\sharp(r)~.
\end{aligned}
\end{equation}
After the change of coordinates, the metric becomes
\begin{align}
\label{btz:inmetric}
\rmd s^2 &= -\left(f-\frac{j^2}{2r^2}\right)\rmd v^2 + 2\rmd v\rmd r - j\rmd v\rmd\psi  + r^2\rmd\psi^2~.
\end{align}
In this system, ingoing light rays follow trajectories with constant $v$ and $\psi$. The outgoing system can be constructed in a similar fashion. 

The causal structure of the BTZ is the same as the RN-AdS case discussed the previous section and depicted in Fig.~\ref{fig:RN.Pd}.

\subsection{Action in a causal diamond}

We now consider a generic causal diamond embedded in the rotating BTZ spacetime, with the same set-up of Fig.~\ref{fig:caus.diamond}, and compute the action. 
%
%

Adopting the outgoing coordinate system $(v, r, \phi)$, the four null boundaries of the diamond are defined by the scalar functions
\begin{subequations}
	\begin{align}
		\label{btz:phi1}
		\Phi_1(v,r) &= v - v_1~,\\
		\label{btz:phi2}
		\Phi_2(v,r) &= u_2 - v + 2r^\ast(r)~,\\
		\label{btz:phi3}
		\Phi_3(v,r) &= v_3 - v~,\\
		\label{btz:phi4}
		\Phi_4(v,r) &= v - 2r^\ast(r) - u_4~.
	\end{align}
\end{subequations}
From these follow the null tangent vectors
\begin{subequations}
	\begin{align}
		\label{btz:k1}
		k_1^\alpha &= \e{\sigma_1} \left( 0, 1 , 0 \right)~,\\
		\label{btz:k2}
		k_2^\alpha &= \e{\sigma_2} \left( \frac{2}{f}, -1, 0 \right)~,\\
		\label{btz:k3}
		k_3^\alpha &= \e{\sigma_3} \left( 0, -1 , 0 \right)~,\\
		\label{btz:k4}
		k_4^\alpha &= \e{\sigma_4} \left( -\frac{2}{f}, 1, 0 \right)~.
	\end{align}
\end{subequations}
From the definition of the non-affinity parameter and \eqref{btz:k1}--\eqref{btz:k4} it follows that $\kappa_i=\partial_\lambda \sigma_i$ for each boundary. With this, the contribution of the four null boundaries is
\begin{align}
	\label{btz:CD.Null}
	I_{N} &= -4\pi\int\limits_{r_3}^{r_2}\rmd r\,\sigma_2 - 4\pi\int\limits_{r_1}^{r_2}\rmd r\,\sigma_1 - 4\pi\int\limits_{r_3}^{r_4}\rmd r\,\sigma_3 - 4\pi\int\limits_{r_1}^{r_4}\rmd r\,\sigma_4 \\
	\notag
	&\quad+ 4\pi r\sigma_2\bigg|_{r_3}^{r_2} + 4\pi r\sigma_1\bigg|_{r_1}^{r_2} + 4\pi r\sigma_3\bigg|_{r_3}^{r_4} + 4\pi r\sigma_4\bigg|_{r_1}^{r_4}~.
\end{align}
The contribution of the four joints is
\begin{align}
\label{btz:CD.joint}
I_{C} &= 4\pi r_1\left[\sigma_1(r_1)+\sigma_4(r_1) - \ln |f(r_1)|\right] - 4\pi r_2\left[\sigma_1(r_2)+\sigma_2(r_2) - \ln |f(r_2)|\right]\\
\notag
&\quad + 4\pi r_3\left[\sigma_2(r_3)+\sigma_3(r_3) - \ln |f(r_3)|\right] - 4\pi r_4\left[\sigma_3(r_4)+\sigma_4(r_4) - \ln |f(r_4)|\right]~.
\end{align}
Finally, using \eqref{btz:metric} to compute the on-shell Lagrangian, the bulk contribution is given by
\begin{align}
	\notag
	I_{B} &= \int\rmd^3x\sqrt{-g}\mathcal{L}\Bigg|_{\text{On-shell}} =  -\frac{8\pi}{L^2}\int\rmd v\rmd r\,r = -\frac{4\pi}{L^2}\int\limits_{v_3}^{v_1}\rmd v\left(\rho_2^2-\rho_4^2\right)\\
	\notag
	&= -4\pi\int\limits_{v_3}^{v_1}\rmd v \left[\rho_2\frac{\rmd\ln |f(\rho_2)|}{\rmd v} + \frac{j^2}{4\rho_2^2} - \left(\rho_4\frac{\rmd\ln |f(\rho_4)|}{\rmd v} + \frac{j^2}{4\rho_4^2}\right)\right]\\
	\label{btz:CD.Bulk}
	&= 4\pi a_v\int\limits_{r_3}^{r_2}\rmd r\,\ln |f(r)| - 4\pi a_v\int\limits_{r_4}^{r_1}\rmd r\,\ln |f(r)| + 4\pi a_u\int\limits_{r_1}^{r_2}\rmd r\,\ln |f(r)| \\
	\notag
	&\quad - 4\pi a_u\int\limits_{r_4}^{r_3}\rmd r\,\ln |f(r)|+ 2\pi j^2\int\limits_{r_4}^{r_1}\frac{\rmd r}{r^2f(r)} - 2\pi j^2\int\limits_{r_3}^{r_2}\frac{\rmd r}{r^2f(r)}\\
	\notag
	&\quad+ 4\pi r_1\ln |f(r_1)| - 4\pi r_4\ln |f(r_4)| + 4\pi r_3\ln |f(r_3)| - 4\pi r_2\ln |f(r_2)|~.
\end{align}
In these manipulations we have used the identity
\begin{equation}
	\label{btz:id}
	\frac{\rho^2}{L^2} = \rho\frac{\rmd\ln |f|}{\rmd v} + \frac{j^2}{4\rho^2}~, 
\end{equation}
and the constants $a_u$ and $a_v$ are again constrained by $a_u+a_v=1$.

By fixing the same type of parameterization used for the previous case, with the parametrization functions proportional to the logarithm of the blackening function 
\begin{equation}
\label{btz:choice}
\begin{aligned}
\text{on $N_1$:}\quad \sigma_1(\lambda) &= a_u \ln |f(r(\lambda))|~, \qquad 
\text{on $N_3$:}\quad \sigma_3(\lambda)  = a_u \ln |f(r(\lambda))|~,\\
\text{on $N_2$:}\quad \sigma_2(\lambda) &= a_v \ln |f(r(\lambda))|~, \qquad 
\text{on $N_4$:}\quad \sigma_4(\lambda)  = a_v \ln |f(r(\lambda))|~,
\end{aligned}
\end{equation}
the total action in the causal diamond is 
\begin{align}
	\label{btz:CD}
	I &= 2\pi j^2\left[\int\limits_{r_4}^{r_1}\frac{\rmd r}{r^2f(r)} - \int\limits_{r_3}^{r_2}\frac{\rmd r}{r^2f(r)}\right]~.	
\end{align}
The parameterization \eqref{btz:choice} is such that the action contribution of any corner vanishes, as it can be seen substituting in \eqref{btz:CD.joint}. Moreover, it satisfies our criterion. This is verified by taking the limit $j\rightarrow 0$ of the action in the causal diamond while keeping fixed the positions $r_i$ of its corners. This limit corresponds to calculating the action in a causal diamond in empty static AdS spacetime, and the limit of \eqref{btz:CD} is zero in accordance to the criterion. 

\subsection{Complexity=Action}

\begin{figure}[t]
	\begin{center}
		\includegraphics[width=.80\textwidth,align=c]{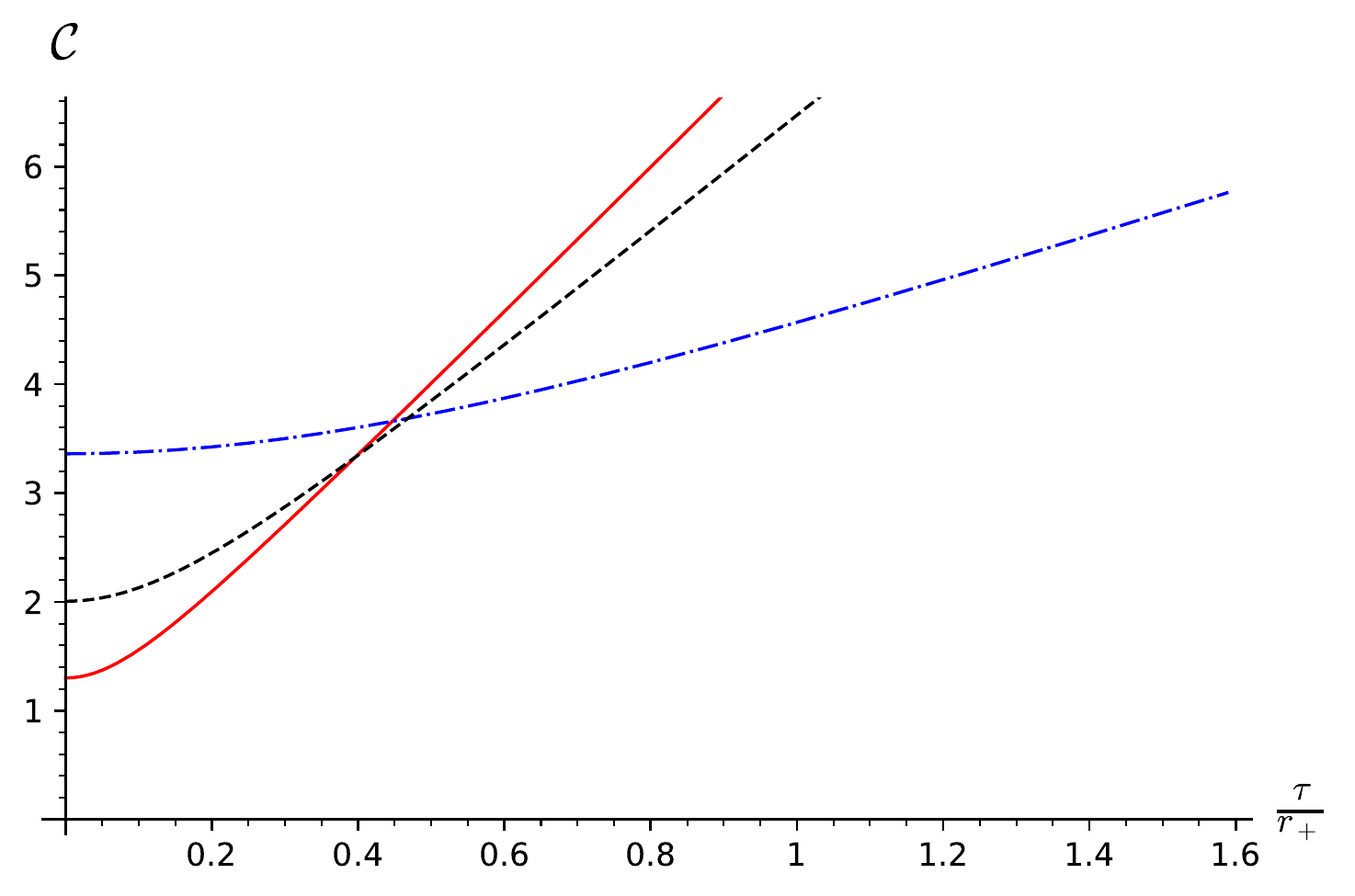}%
		\hfill%
		\caption{Time evolution of the complexity of the rotating BTZ black hole for different values of the angular momentum. Solid line: $J = r_+$, dash-dotted line: $J = 1.5r_+$, dashed line: $J = 2.2r_+$. ($L = 0.8r_+$, $8G = \hbar = 1$ and $m=3$ for all plots).\label{fig:C.BTZ}}
	\end{center}
\end{figure}

\begin{figure}[t]
	\begin{center}
		\includegraphics[width=.75\textwidth]{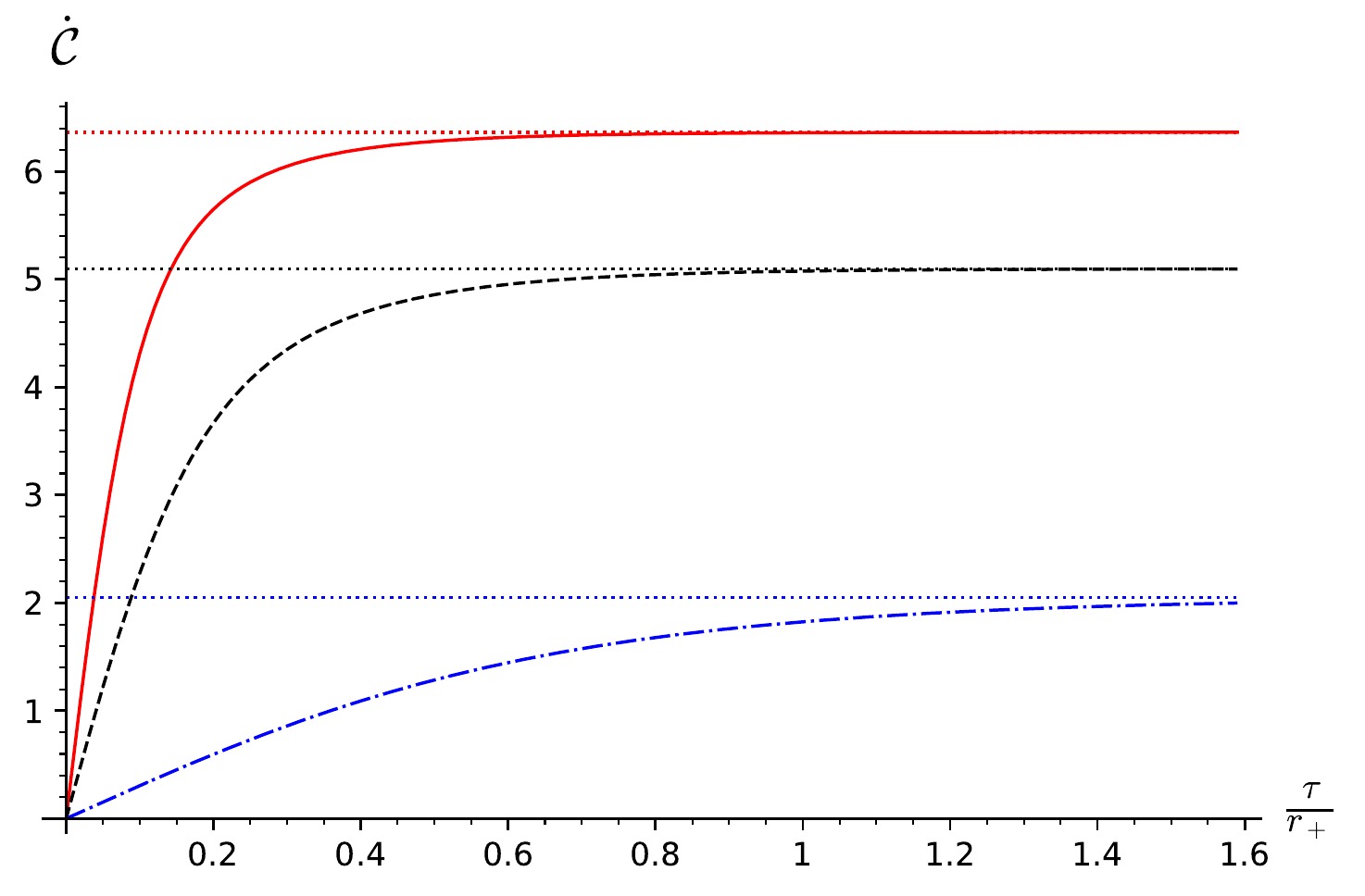}%
		\hfill%
		\caption{Growth rate of the complexity of the rotating BTZ black hole for different values of the angular momentum. Solid line: $J = r_+$, dash-dotted line: $J = 1.5r_+$, dashed line: $J = 2.2r_+$. For each value of $J$ Lloyd's bound is represented by the dotted horizontal line. ($L = 0.8r_+$, $8G = \hbar = 1$ and $m=3$ for all plots).\label{fig:Cdot.BTZ}}
	\end{center}
\end{figure}

The WdW patch of the rotating BTZ black hole, just like in the previous RN-AdS case, is a causal diamond. The null boundaries of the patch meet at radii $r_m^\pm$, defined by $r^\ast(r_m^\pm) = \pm\tau$.  When $\tau = 0$ we have $r_m^- = r_m^+$ and, for $\tau\to \infty$, $r_m^\pm\to r_\mp$. Therefore, from \eqref{btz:CD} and \eqref{intro:complexity}, the complexity is
\begin{equation}
	\label{btz:I.WdW}
	\mathcal{C} = -\frac{8GJ^2}{\pi\hbar}\left[\int\limits_{r_m^+}^{r_m^-}\frac{\rmd r}{r^2f(r)} + 2\int\limits_{r_m^-}^{R}\frac{\rmd r}{r^2f(r)}\right]
\end{equation}
The second integral turns out to be finite for $R\to \infty$, implying that the complexity is again finite. Moreover, in this simple case we are able to carry out the integration and have an expression in closed form,
\begin{equation}
	\label{btz:C}
	\mathcal{C}=\frac{JL}{2\pi\hbar\sqrt{\frac{M^2L^2}{J^2} - 1}}\left[\frac{1}{r_+}\ln\frac{(r_+ - r_m^+)(r_+ - r_m^-)}{(r_+ + r_m^+)(r_+ + r_m^-)} - \frac{1}{r_-}\ln\frac{(r_m^+ - r_-)(r_m^- - r_-)}{(r_m^+ + r_-)(r_m^- + r_-)}\right]~.
\end{equation}
By plotting \eqref{btz:C} as a function of boundary time $\tau$, as in Fig.~\ref{fig:C.BTZ}, we see that the complexity is always positive and increasing, reaching a linear growth regime at late times.
The fact that the complexity is always increasing is also confirmed by computing its time derivative
\begin{equation}
	\label{btz:I.der}
	\frac{\rmd{\mathcal{C}}}{\rmd\tau} = \frac{8GJ^2}{\pi\hbar}\left[\frac{1}{(r_m^+)^2} - \frac{1}{(r_m^-)^2}\right]~,
\end{equation}
which, because $r_m^+\leq r_m^-$ for $\tau\geq0$, is always non-negative. A plot of the complexity growth rate \eqref{btz:I.der} is shown in Fig.~\ref{fig:Cdot.BTZ}. The complexity growth rate is zero at $\tau = 0$ and then increases monotonically. At late times, \eqref{btz:I.der} reaches the asymptotic value
\begin{equation}
\label{btz:I.derlimit}
\lim_{\tau\rightarrow\infty}\frac{\rmd\mathcal{C}}{\rmd\tau} = \frac{8GJ^2}{\pi\hbar}\left(\frac{1}{r_-^2} - \frac{1}{r_+^2} \right) = \frac{4}{\pi\hbar}\sqrt{M^2-\frac{J^2}{L^2}},
\end{equation}
which agrees with previous results \cite{Brown:2015lvg, Carmi:2017jqz, Cai:2016xho} and exactly saturates the upper bound proposed in \cite{Cai:2016xho}\footnote{For the rotating BTZ black hole, this bound and the one proposed in \cite{Brown:2015lvg} are equivalent}. In this case, the monotonic increase of \eqref{btz:I.der} can be proved computing the second derivative of the complexity
\begin{equation}
	\frac{\rmd^2\mathcal{C}}{\rmd\tau^2}= -\frac{16GJ^2}{\pi\hbar}\left(\frac{f(r_m^+)}{(r_m^+)^3} + \frac{f(r_m^-)}{(r_m^-)^3}\right)~,
\end{equation}
which is always positive because $f(r)<0$ when $r_-<r<r_+$. This shows how the (generalized) Lloyd bound is never violated during the time evolution of the black hole.

\begin{figure}[t]
	\begin{center}
		\includegraphics[width=.75\textwidth]{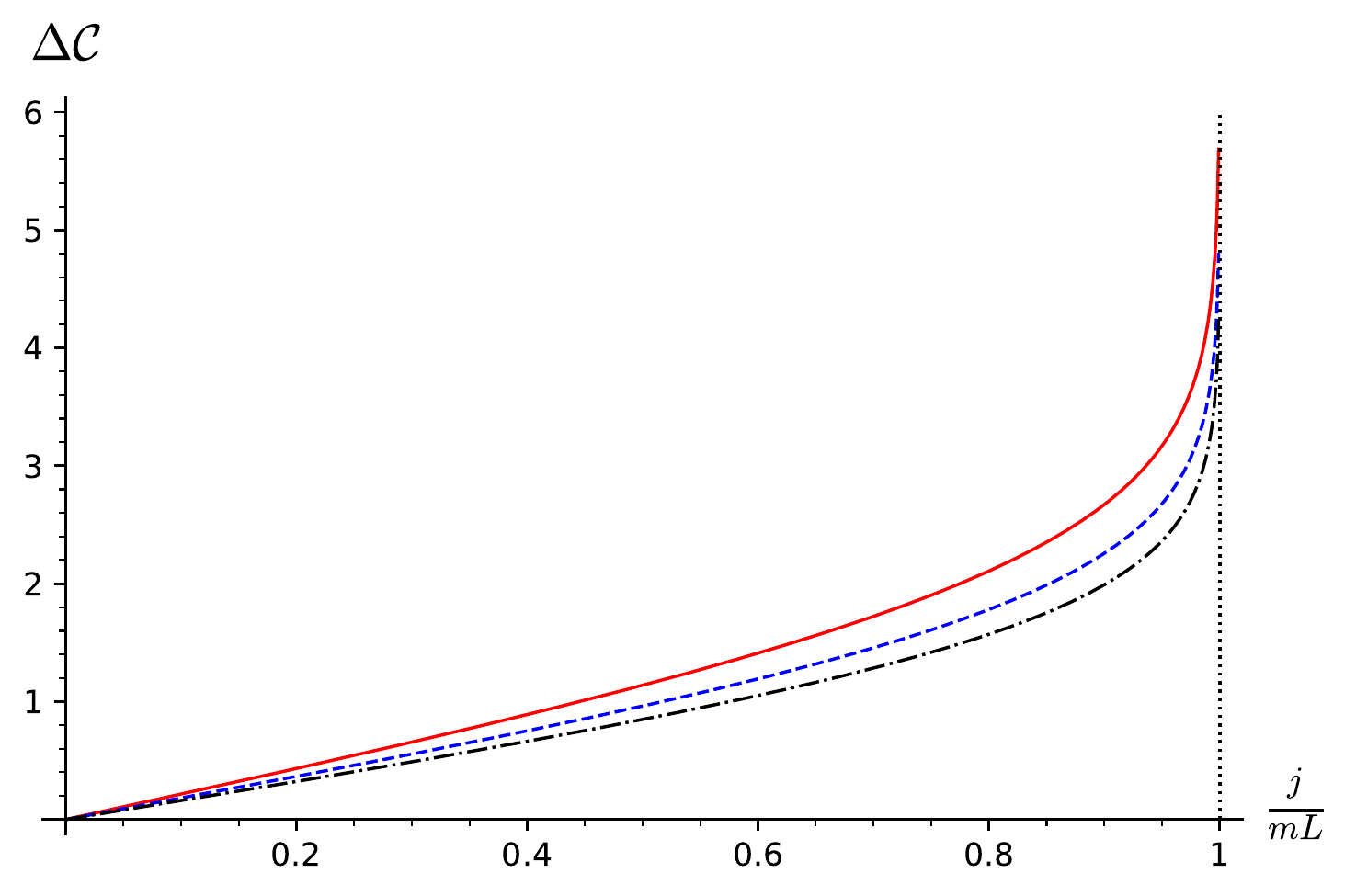}%
		\hfill%
		\caption{Complexity of formation of the rotating BTZ black hole for different values of the mass parameter $m$. Solid line: $m = 5$, dashed line: $m = 7$, dashed-dotted line: $m = 9$ ($8G = \hbar = 1$ for all plots).\label{fig:BTZ_formation}}
	\end{center}
\end{figure}

The simple expression for the complexity \eqref{btz:I.WdW} allows us to discuss the complexity of formation of the rotating BTZ black hole. The complexity of formation is defined by the difference between the complexity of the black hole at initial time $\tau = 0$ and the complexity of the asymptotic geometry in which it has formed, in our case pure AdS$_3$ spacetime. In our approach, AdS$_3$ has vanishing complexity, so that the complexity of formation is given by \eqref{btz:C} evaluated at $\tau = 0$. The expression of the complexity of formation is then
\begin{equation}
	\Delta\mathcal{C} = \frac{JL}{\pi\hbar\sqrt{\frac{M^2L^2}{J^2} - 1}}\left[\frac{1}{r_+}\ln\frac{(r_+ - r_m^0)}{(r_+ + r_m^0)} - \frac{1}{r_-}\ln\frac{(r_m^0 - r_-)}{(r_m^0 + r_-)}\right]~,
\end{equation}
where we have defined $r_m^0 = r_m^+(0) = r_m^-(0)$. The value of $r_m^0$ can be obtained numerically solving $r^\ast(r_m^0)=0$. A plot of the complexity of formation as a function of the dimensionless quantity $\frac{j}{mL}$ is shown in Fig.\ref{fig:BTZ_formation}. As the angular momentum parameter $j$ varies in the permitted range $[0, mL]$, the dimensionless ratio varies between $0$, which corresponds to a static BTZ black hole, and $1$, which corresponds to an extremal rotating BTZ black hole. The plots show that the complexity of formation is zero for $J = 0$, which is just the result for the static BTZ black hole we found in \cite{Mounim:2021bba}, and then monotonically increases with the angular momentum of the black hole. This is in contrast with the findings of \cite{Bernamonti:2021jyu}, where the complexity of formation is computed in two ways, with the inclusion of the counterterm which renders the action reparameterization invariant, and using the affine parameterization. In both cases they find that the complexity of formation is negative and decreasing for small angular momentum of the black hole, and eventually starts to increase and becomes positive for big enough angular momentum. In the extremal black hole limit, when $\frac{j}{mL}\rightarrow 1$, or equivalently $r_+\rightarrow r_-$, the complexity of formation diverges as
\begin{equation}
	\Delta\mathcal{C} \sim \frac{1}{\pi\hbar\sqrt{m}}\ln\frac{2r_+}{r_+-r_-}~.
\end{equation}
This logarithmic divergence is also observed in \cite{Bernamonti:2021jyu}, although with a different leading coefficient. In particular, the coefficient of divergence we have found is proportional to the inverse of the coefficient found by \cite{Bernamonti:2021jyu}.
\section{Kerr-AdS black hole}
\label{kads}

\subsection{Setup}

Another interesting and more complicated example of a vacuum solution of Einstein gravity with a negative cosmological constant is the Kerr-AdS black hole, which is non-static. This solution describes a rotating black hole in an asymptotically AdS spacetime with more than three dimensions. For simplicity, we consider a Kerr-Ads black hole in four dimensions. Its metric is \cite{Carter:1968ks}
\begin{align}
\label{kads:metric}
\rmd s^2 &= -\frac{\Delta_r - a^2\Delta_\theta\sin^2\theta}{\Sigma^2}\rmd t^2 + \frac{\Sigma^2}{\Delta_r}\rmd r^2 + \frac{\Sigma^2}{\Delta_\theta}\rmd \theta^2 + 2a\sin^2\theta\frac{\Delta_r - (r^2+a^2)\Delta_\theta}{\Sigma^2\Xi}\rmd t\rmd\phi\\
\notag
&\quad + \frac{\Pi\sin^2\theta}{\Sigma^2\Xi^2}\rmd\phi^2~,
\end{align}
where\footnote{Notice that our notation differs from \cite{Bernamonti:2021jyu}. In particular, our $\Sigma^2$ corresponds to their $\rho^2$, and our $\Pi$ to their $\Sigma^2$.} 
\begin{equation}
\label{kads:defs}
\begin{aligned}
	\Delta_r &= (r^2+a^2)\left(1+\frac{r^2}{L^2}\right) - 2mr~, \qquad 
	\Delta_\theta = 1-\frac{a^2\cos^2\theta}{L^2}~, \\
	\Sigma^2 &= r^2 + a^2\cos^2\theta~, \qquad \Xi = 1-\frac{a^2}{L^2}~, \qquad 
	\Pi = \left(r^2+a^2\right)^2\Delta_\theta-a^2\Delta_r\sin^2\theta~.
\end{aligned}
\end{equation}
The parameters $m>0$ and $0<a<L$ determine the physical mass and angular momentum of the black hole
\begin{equation}
\label{kads:MJ}
M = \frac{m}{G\Xi^2}, \qquad 
J = \frac{am}{G\Xi^2}~.
\end{equation}

For the zeroes of $\Delta_r$ to be positive and distinct, the mass parameter must satisfy\cite{Bernamonti:2021jyu}
\begin{equation}
	\label{kads:masscondition}
	m>\frac{L}{3\sqrt{6}}\left(\sqrt{\left(1+\frac{a^2}{L^2}\right)^2+\frac{12a^2}{L^2}}+\frac{2a^2}{L^2}+2\right)\sqrt{\sqrt{\left(1+\frac{a^2}{L^2}\right)^2+\frac{12a^2}{L^2}}-\frac{a^2}{L^2}-1}~.
\end{equation}
We will assume \eqref{kads:masscondition} for the rest of the section. It is, actually, simpler to rewrite $\Delta_r$ as 
\begin{equation}
\label{kads:Delta}
	\Delta_r = \frac1{L^2} (r-r_+)(r-r_-)\left[r^2 + (r_-+r_+)r +\alpha \right]~,
\end{equation}
where $\alpha$, $r_-$ and $r_+$ satisfy the relations
\begin{align}
\label{kads:alpha1}
	\alpha r_- r_+ &= a^2 L^2~,\\
\label{kads:alpha2}
	(r_+ + r_-)(\alpha -r_-r_+) &= 2m L^2~,\\
\label{kads:alpha3}
	\alpha +r_-r_+ - (r_+ + r_-)^2 &= a^2 +L^2~, 
\end{align}
subject to the condition\footnote{The condition \eqref{kads:condition} unifies the two conditions $a<L$ and $r_+^2>aL$ necessary to avoid  super-luminal rotation velocities at the outer horizon.} 
\begin{equation}
\label{kads:condition}
	\frac{a}{r_+} < \min \left( \frac{L}{r_+}, \frac{r_+}{L} \right)~.
\end{equation}

The geometry defined by \eqref{kads:metric} has a curvature singularity with the topology of a ring at $r=0$ and $\theta = \frac{\pi}{2}$, as well as two horizons at $r_\pm$, defined by $\Delta_r(r_\pm) = 0$. As shown in \cite{Balushi:2019pvr}, the geometry is free of caustics between the horizons and outside of the outer horizon. 

\begin{figure}[t]
	\hfill
	\begin{subfigure}{.45\textwidth}
		\centering
		\includegraphics[height=0.4\textheight]{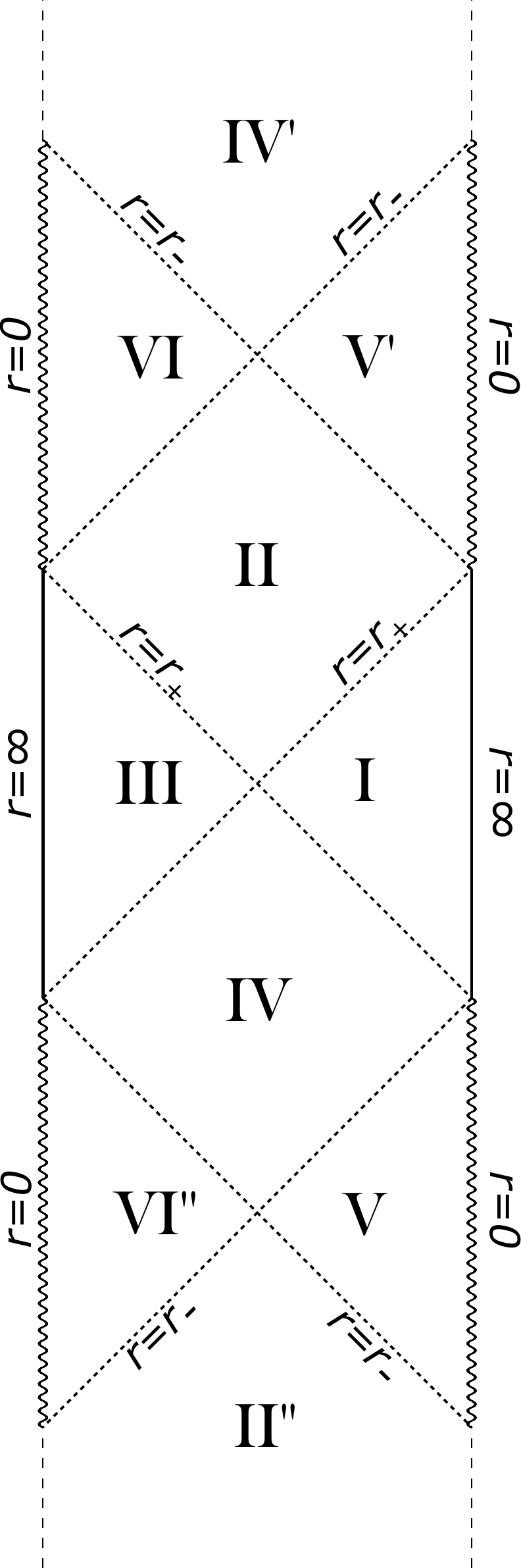}
		\caption{$\theta = \frac{\pi}{2}$}
	\end{subfigure}%
	\hfill
	\begin{subfigure}{.45\textwidth}
		\centering
		\includegraphics[height=0.4\textheight]{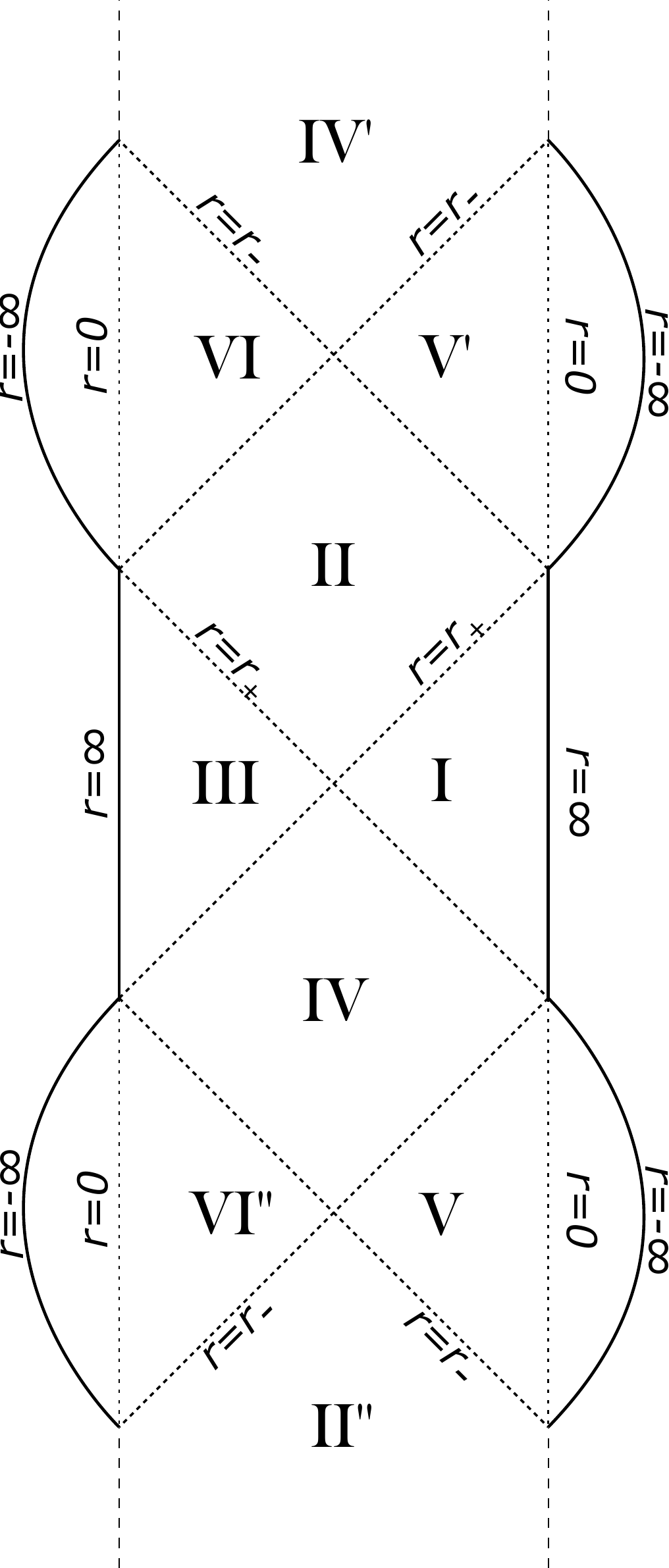}
		\caption{$\theta\neq\frac{\pi}{2}$}
	\end{subfigure}
	\hfill
	\caption{The Penrose diagram of Kerr-AdS spacetime. (a) The singularity is located at $r=0$, $\theta=\frac{\pi}2$. (b) For $\theta\neq \frac{\pi}2$, the spacetime can be continued to $r<0$.}
	\label{fig:KerrPenrose}
\end{figure}

The Penrose diagram of Kerr-AdS spacetime depends on whether or not $\theta=\frac{\pi}{2}$, because of the location of the curvature singularity mentioned above. The two possible Penrose diagrams are shown in Fig.~\ref{fig:KerrPenrose}. When $\theta=\frac{\pi}{2}$, which includes the curvature singularity at $r=0$, the Penrose diagram is similar to the diagram of RN-AdS spacetime. If $\theta\neq\frac{\pi}{2}$, $r=0$ is a regular point, and the spacetime can be analytically continued to a region with $r<0$. In any case, this distinction is irrelevant for our purposes, because the WdW patch is located entirely in the regions I--IV. 

The relevant thermodynamic quantities are associated with the outer horizon. They are 
\begin{equation}
\label{kads:thermo}
\Omega_+ = \frac{a}{L^2}\frac{r_+^2+L^2}{r^2_++a^2}~,\quad 
T = \frac{r_+}{2\pi L^2}\frac{r_+^2+L^2}{r_+^2+a^2}-\frac{1}{4\pi r_+}\left(1-\frac{r_+^2}{L^2}\right)~,\quad 
S = \frac{\pi r_+^2}{4G},
\end{equation}
which are the horizon angular velocity, the temperature, and the entropy, respectively.

To study the black hole, we introduce ingoing and outgoing Kerr null coordinates
\begin{equation}
\label{kads:ingoing}
\begin{aligned}
&v = t + r^\ast(r, \theta)~,\\
&u = t - r^\ast(r, \theta),
\end{aligned}
\end{equation}
where the tortoise coordinate $r^\ast$ is now dependent also on the angular variable $\theta$. To be specific, let us consider a $v = \const$ null hypersurface. The tangent covector to such a hypersurface, which we shall call $l$, is given by 
\begin{equation}
\label{kads:ldef}
	l_\mu = \partial_\mu v = \left(1, \partial_r r^\ast, \partial_\theta r^\ast, 0\right)~,
\end{equation}
Its norm ought to be zero, which yields a differential equation for $r^\ast$,
\begin{equation}
	\label{kads:diffeq}
	\Delta_r\left(\partial_r r^\ast\right)^2 + \Delta_\theta\left(\partial_\theta r^\ast\right)^2 = \frac{\left(r^2+a^2\right)^2}{\Delta_r} - \frac{a^2\sin^2\theta}{\Delta_\theta}~.
\end{equation}
The solution to \eqref{kads:diffeq} can be expressed as follows \cite{Balushi:2019pvr, Pretorius:1998sf}
\begin{align}
	\label{kads:PandQ}
	\partial_r r^\ast(r, \theta) &= \frac{Q}{\Delta_r},\\
	\partial_\theta r^\ast(r, \theta) &= \frac{P}{\Delta_\theta},
\end{align}
where
\begin{equation}
	\label{kads:PQ}
	\begin{aligned}
		Q^2 &= (r^2 + a^2)^2 - a^2\zeta\Delta_r,\\
		P^2 &= a^2\left(\zeta\Delta_\theta - \sin^2\theta\right)~,
	\end{aligned}
\end{equation}
and $\zeta(r,\theta)$ is an auxiliary function. It must obey a consistency condition imposed by $\rmd^2 r^\ast = 0$, which reads
\begin{equation}
	\label{kads:dzeta}
	\rmd\zeta = \frac{1}{\mu}\left(P\rmd r - Q\rmd\theta\right)~,
\end{equation}
where $\mu(r,\theta)$ is an integrating factor again subject to the condition $\rmd^2\zeta = 0$. 

Notice that from \eqref{kads:PQ} and \eqref{kads:defs} follows
\begin{equation}
	\Pi = \Delta_\theta Q^2 + \Delta_r P^2~.
\end{equation}

Similar considerations hold for the outgoing coordinate $u$. The tangent covector to a $u = \const$ hypersurface, which we call $n$, reads
\begin{equation}
	n_\mu = \partial_\mu u = \left(1, -\partial_r r^\ast, -\partial_\theta r^\ast, 0\right)~.
\end{equation}
It is null when $r^\ast$ satisfies again \eqref{kads:diffeq}.

Because $\zeta$ satisfies $l^\mu\partial_\mu\zeta = n^\mu\partial_\mu\zeta = 0$, it is convenient to adopt it as an intrinsic coordinate along a space-like direction, together with $\phi$. The space-like tangents are 
\begin{align}
	e^\mu_\zeta &= \frac{\partial x^\mu}{\partial\zeta} = \frac{\mu}{\Pi} \left( 0,P\Delta_r, -Q \Delta_\theta, 0\right)~,\\
	e^\mu_\phi &= \frac{\partial x^\mu}{\partial\phi} =  \left( 0,0, 0, 1\right)~,
\end{align}
which determine the non-degenerate induced metric 
\begin{equation}
	\label{kads:indMetric}
	\gamma_{ab} = \begin{pmatrix}
		\frac{\mu^2\Sigma^2}{\Pi} & 0\\
		0 & \frac{\Pi\sin^2\theta}{\Sigma^2\Xi^2}
	\end{pmatrix}~.
\end{equation} 

The null coordinate along a $v=\text{const.}$ hypersurface is determined implicitly by 
\begin{equation}
\label{kads:k.def}
	k^\mu = \frac{\partial x^\mu}{\partial\lambda} = \e{\sigma(\lambda,\zeta)} l^\mu~, 
\end{equation} 
and similarly for a $u=\text{const.}$ hypersurface with $n^\mu$ instead of $l^\mu$. The function $\sigma(\lambda,\zeta)$ implements the freedom of parameterization. By symmetry, we take $\sigma$ to be independent of $\phi$.

\subsection{Action in a causal diamond}

In this subsection, we will discuss the action in a generic causal diamond in Kerr-AdS spacetime. The setup is of the causal diamond is the same as in the RN-AdS case, see Fig.~\ref{fig:caus.diamond}. We shall consider the diamond in the coordinates $(u,r,\theta,\phi)$, with $u_2\leq u\leq u_4$ and 
$\rho_3(u,\theta) \leq r \leq \rho_1(u,\theta)$, where $\rho_1$ and $\rho_3$ are implicitly determined by the defining relations of the null boundaries $N_1$ and $N_3$, 
\begin{equation}
\label{kads:rho.13}
	u + 2 r^\ast (\rho_1,\theta ) = v_1~,\qquad u + 2 r^\ast (\rho_3,\theta ) = v_3~.
\end{equation}

First, we consider the contribution of the four joints. Taking the first joint as an example, the null tangent vectors that meet in the joint are $k_1^\mu = \e{\sigma_1}l^\mu$ and $k_4^\mu = \e{\sigma_4}n^\mu$. Therefore, the contribution of this joint reads
\begin{align}
	\label{kads:I.C14}
	I_{C_{41}} &= 2\int\rmd\phi\rmd\zeta\sqrt{\gamma}\ln\left|\frac{k_1\cdot k_4}{2}\right| = \frac{4\pi}{\Xi}\int\rmd\zeta\left[|\mu|\sin\theta\ln\left|\frac{\Pi\e{\sigma_1 + \sigma_4}}{\Sigma^2\Delta_r\Delta_\theta}\right|\right]_{C_{41}}~.
\end{align}
The other three joints yield analogous contributions.

We proceed fixing the parameterization functions $\sigma_1,\ldots, \sigma_4$. Motivated by analogy with the simpler cases of the Schwarzschild-AdS black hole \cite{Mounim:2021bba}, the RN-AdS black hole and the rotating BTZ black hole, we shall choose a parameterization, in which the corner terms vanish. For simplicity, let us take 
\begin{equation}
	\label{kads:choice}
		\sigma_1(\lambda,\zeta) = \ln\left|\frac{\Sigma^2\Delta_r\Delta_\theta}{\Pi}\right|_{N_1}~, \quad
		\sigma_3(\lambda,\zeta) = \ln\left|\frac{\Sigma^2\Delta_r\Delta_\theta}{\Pi}\right|_{N_3}~,\quad
		\sigma_2(\lambda,\zeta) = \sigma_4(\lambda,\zeta) =0~,
\end{equation}
which corresponds to $a_u=1$ and $a_v=0$ in the previous cases. The suffixes $N_1$ and $N_3$ indicate pull-backs to the null surfaces.

Next, we consider the contributions of the null boundaries. For a $v=\text{const.}$ boundary ($N_1$ or $N_3$), the tangent vector $k^\mu$ is given by \eqref{kads:k.def}, $k^\mu =\e{\sigma} l^\mu$. Because $l^\nu\nabla_\nu l^\mu = 0$ from \eqref{kads:ldef}, one easily computes
\begin{equation}
	\label{kads:kappa}
	k^\nu\nabla_\nu k^\mu = (\partial_\lambda\sigma)  k^\mu~,
\end{equation}
so that the non-affinity parameter is $\kappa= \partial_\lambda \sigma$. Hence, for $N_1$, we have the contribution
\begin{align}
	\label{kads:I.N1}
	I_{N_1} &= \frac{4\pi}{\Xi}\int\rmd\lambda\rmd\zeta\,|\mu|\sin\theta\,\partial_\lambda\sigma_1~.
\end{align}
An analogous contribution arises from $N_3$. Instead, $N_2$ and $N_4$ do not contribute, because of our choice \eqref{kads:choice}. 

It is convenient to rewrite \eqref{kads:I.N1} by changing coordinates from $(\lambda,\zeta)$ to $(u, \theta)$. The Jacobian of the coordinate change is
\begin{equation}
	\frac{\partial (u,\theta)}{\partial(\lambda, \zeta)} = \begin{pmatrix}
			-\frac{2\Pi\e{\sigma_1}}{\Sigma^2\Delta_r\Delta_\theta} & 0\\
			\frac{P\e{\sigma_1}}{\Sigma^2} & -\frac{\mu\Delta_\theta Q}{\Pi}
	\end{pmatrix}~.
\end{equation}
We observe in passing that the choice \eqref{kads:choice} implies $u=\mp 2\lambda +\text{const.}$, where the sign depends on the sign of  $\Delta_r$. After the change of coordinates, \eqref{kads:I.N1} becomes
\begin{equation}
		\label{kads:I.N1utheta}
		I_{N_1} = \frac{2\pi}{\Xi}\int\limits_{0}^{\pi}\rmd\theta\sin\theta\int\limits_{u_2}^{u_4}\rmd u\left[\frac{\Delta_r\Sigma^2}{Q} l^\mu \partial_\mu \sigma_1\right]^{r=\rho_1}~.
\end{equation}
The choice \eqref{kads:choice} implies that $\sigma$ depends only on the spacetime coordinates $r$ and $\theta$. After calculating also $l^\mu$ by raising the index in \eqref{kads:ldef}, \eqref{kads:I.N1utheta} reads   
\begin{equation}
		\label{kads:I.N1utheta2}
		I_{N_1} = \frac{2\pi}{\Xi}\int\limits_{0}^{\pi}\rmd\theta\sin\theta\int\limits_{u_2}^{u_4}\rmd u\left[\Delta_r \left( \partial_r +\frac{P}{Q} \partial_\theta\right) \sigma_1\right]^{r=\rho_1}~.
\end{equation}
Finally, substituting \eqref{kads:choice} and adding also the contribution from $N_3$, the action contribution from the null boundaries is found to be
\begin{align}
	\label{kads:I.N_2}
	I_{N} = I_{N_1} + I_{N_3} = \frac{2\pi}{\Xi}\int\limits_{0}^{\pi}\rmd\theta\sin\theta\int\limits_{u_2}^{u_4}\rmd u\left[\Delta_r \left(\partial_r + \frac{P}{Q}\partial_{\theta}\right) \ln\left|\frac{\Sigma^2\Delta_r\Delta_\theta}{\Pi}\right|\right]^{r=\rho_1}_{r=\rho_3}~.
\end{align}

The last ingredient is the bulk action, which is 
\begin{align}
	\notag
	I_B &= -\frac{6}{L^2}\int\rmd u\rmd\theta\rmd\phi\rmd r\frac{\Sigma^2\sin\theta}{\Xi} = -\frac{12\pi}{\Xi L^2}\int\rmd u\rmd \theta\sin\theta \rmd r\,\Sigma^2\\
	\label{kads:I.B}
	&= -\frac{4\pi}{\Xi L^2}\int\limits_{0}^{\pi}\rmd\theta\sin\theta\int\limits_{u_2}^{u_4}\rmd u\Big[r^3 + 3a^2r\cos^2\theta\Big]_{r=\rho_3}^{r=\rho_1}~.
\end{align}

Summing the contributions \eqref{kads:I.N_2} and \eqref{kads:I.B}, the total action of the causal diamond is given by
\begin{equation}
	\label{kads:I.CD}
	I =\frac{4\pi}{\Xi}\int\limits_{0}^{\pi}\rmd\theta\sin\theta\int\limits_{u_2}^{u_4}\rmd u\,\Big[F(\rho_3, \theta)-F(\rho_1,\theta)\Big]~,
\end{equation}
where we have introduced the function $F(r,\theta)$ by 
\begin{equation}
	\label{kads:Fold}
	F(r,\theta) = -\frac{r^3 + 3a^2r\cos^2\theta}{L^2} + \frac{r^2}{2} \partial_r \left(\frac{\Delta_r}{r^2}\right) 
	-\frac{\Delta_r}{2} \left(\partial_r +\frac{P}{Q} \partial_\theta\right) \ln \left|\frac{\Pi}{r^2\Sigma^2 \Delta_\theta}\right|~.
\end{equation}
This can be slightly simplified by noting that
\begin{equation}
\label{kads:delr.Delta}
	\partial_r \left(\frac{\Delta_r}{r^2}\right) =2 \left( \frac{r}{L^2} - \frac{a^2}{r^3} +\frac{m}{r^2} \right)~, 
\end{equation} 
which gives
\begin{equation}
\label{kads:F}
	F(r,\theta) = -\frac{3a^2r\cos^2\theta}{L^2} +m -\frac{a^2}r 
	-\frac{\Delta_r}{2} \left(\partial_r +\frac{P}{Q} \partial_\theta\right) \ln \left|\frac{\Pi}{r^2\Sigma^2 \Delta_\theta}\right|~.
\end{equation}
The term $m$ can be omitted in this expression, because it cancels in \eqref{kads:I.CD}. We also have the following identity, 
\begin{equation}
\label{kads:Pi.ident}
	\frac{\Pi}{r^2\Sigma^2} =\Xi\left(1+\frac{a^2}{r^2} +\frac{2ma^2\sin^2\theta}{r\Sigma^2 \Xi} \right)~.
\end{equation} 
This immediately shows that \eqref{kads:I.CD} vanishes for $a=0$, in agreement with the result that the action in any causal diamond of Schwarzschild-AdS vanishes \cite{Mounim:2021bba}.

For the following, it will be convenient to express the result \eqref{kads:I.CD} using $r$ as the integration variable instead of $u$. Taking into account the above comment on $m$, we get
\begin{align}
	\label{kads:I.CDr}
	I &= \frac{8\pi}{\Xi}\int\limits_{0}^{\pi}\rmd\theta\sin\theta\left[
	\int\limits_{r_1(\theta)}^{r_2(\theta)}\rmd r\frac{Q(F-m)}{\Delta_r} + \int\limits_{r_3(\theta)}^{r_4(\theta)}\rmd r\frac{Q(F-m)}{\Delta_r}\right]~.
\end{align}

\subsection{Complexity=Action}

\begin{figure}[t]
	\centering
	\includegraphics[width=0.3\textwidth]{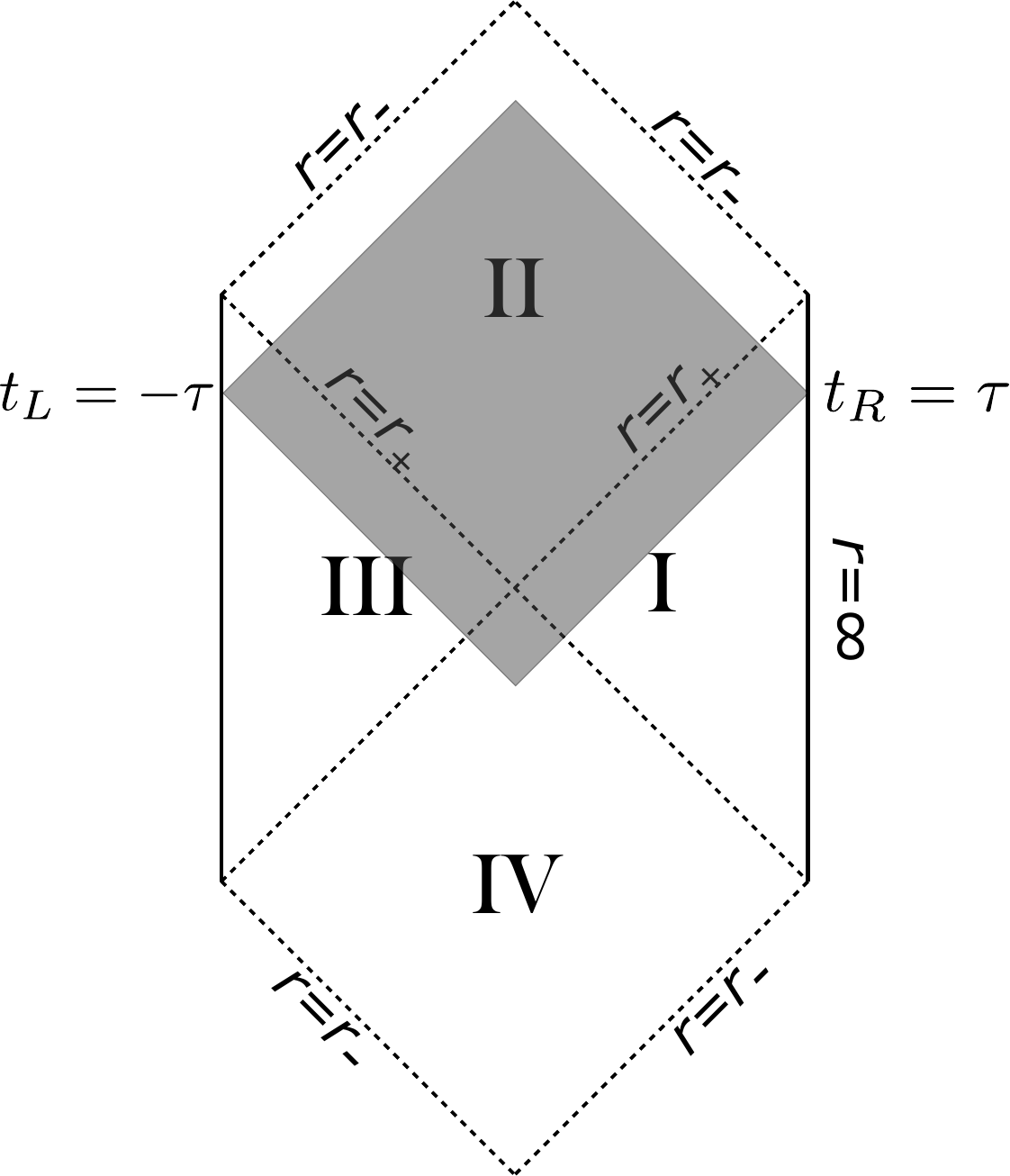}
	\caption{WdW patch in Kerr-AdS spacetime.\label{fig:kads.WdW}}
\end{figure}

The WdW patch of the Kerr-AdS black hole is a causal diamond anchored at two cut-off boundaries located at $r = R\gg L$. It is located entirely in the quadrants I--IV, see Fig.~\ref{fig:kads.WdW}. As before, we use time translation invariance to fix $t=\tau$ and $t=-\tau$ on the right and left boundaries, respectively. The null boundaries of the WdW patch meet in the quadrants II and IV at the locations $r=r_m^+$ and $r=r_m^-$, respectively. These satisfy 
\begin{equation}
	\label{kads:rmeet}
	r^\ast(r_m^\pm, \theta) = \pm\tau + r^\ast(R,\theta)~,
\end{equation}
together with $r_-<r_m^{\pm}<r_+$. For $\tau>0$, we have $r_m^+<r_m^-$ ($r_m^+=r_m^-$ for $\tau=0$), and $r_m^\pm \to r_\mp$ for $\tau\to \infty$.

From \eqref{kads:rmeet} and \eqref{kads:PandQ} one obtains the derivative of $r_m^\pm$ with respect to $\tau$,
\begin{align}
	\partial_\tau r_m^\pm = \pm\frac{\Delta_r}{Q}\Bigg|_{r= r_m^\pm}~.
\end{align} 

We are now ready to write down the complexity. Applying \eqref{kads:I.CDr} to the WdW patch and translating to complexity, we get 
\begin{align}
	\label{kads:C}
	\mathcal{C} = \frac{1}{2\pi\hbar\Xi G}\int\limits_{0}^{\pi}\rmd\theta\sin\theta\left[\int\limits_{r_m^+}^{R}\rmd r\frac{Q}{\Delta_r}(F-m) + \int\limits_{r_m^-}^{R}\rmd r\frac{Q}{\Delta_r}(F-m)\right]~.
\end{align} 
The integrands of \eqref{kads:C} have a pole at $r = r_+$. This is, however, an integrable singularity. 

It turns out that the integrals do not diverge for large $R$, so that the cut-off can be removed. 
To show this, it is necessary to obtain $P$ and $Q$ for large $r$. For the auxiliary variable $\zeta$ in \eqref{kads:PQ}, 
let us use the ansatz
\begin{equation}
\label{kads:zeta.ans}
	\zeta = \frac{\sin^2\theta}{\Delta_\theta} + \frac{\alpha^2(\theta)}{r^2} + \cdots
\end{equation}
with an undetermined function $\alpha(\theta)$. Here and henceforth, the ellipses denote irrelevant sub-leading terms in $r$. Together with \eqref{kads:PQ}, \eqref{kads:zeta.ans} implies 
\begin{equation}
\label{kads:PandQ.r}
	Q = r^2 \sqrt{\frac{\Xi}{\Delta_\theta}}+\cdots~,\qquad 
	P = \frac{a}{r}\alpha(\theta) \sqrt{\Delta_\theta}+ \cdots~.
\end{equation}
Furthermore, \eqref{kads:dzeta} gives 
\begin{equation}
\label{kads:PandQ.mu.r}
	\frac{Q}{\mu} = -2 \sin\theta \cos\theta \frac{\Xi}{\Delta_\theta^2}+\cdots~, \qquad  
	\frac{P}{\mu} = -\frac{2\alpha^2(\theta)}{r^3}+\cdots~.
\end{equation}
From \eqref{kads:PandQ.mu.r} and \eqref{kads:PandQ.r} one can determine $\alpha(\theta)$ and, subsequently, 
\begin{equation}
\label{kads:P.r}
	P = \frac{a^2}r \sqrt{\frac{\Xi}{\Delta_\theta}} \sin\theta\cos\theta +\cdots~.
\end{equation}

Returning to the complexity \eqref{kads:C}, the dominant term in the integrand goes like $1/r$ for large $r$ and has three contributions, which correspond to the terms that grow as $r$ in \eqref{kads:F}. The first contribution is simply the first term on the right hand side, the second one arises from the term with $\partial_r$ upon using \eqref{kads:Pi.ident}, and the last contribution is $\frac{P}{2\Delta_\theta} \partial_\theta\Delta_\theta$. Summing up these three contributions, one gets
\begin{align}
\notag
	\frac{Q}{\Delta_r} (F-m) &= \frac{a^2}{r} \sqrt{\frac{\Xi}{\Delta_\theta}}
	\left[-3\cos^2\theta + 1 +\frac{a^2 \sin^2\theta\cos^2\theta}{L^2\Delta_\theta}\right] +\cdots \\
\label{kads:integrand.large.r}
	&=\frac{a^2}{r} \sqrt{\frac{\Xi}{\Delta_\theta^3}}
	\left(1-3\cos^2\theta +2 \frac{a^2}{L^2} \cos^4 \theta \right)+\cdots~.
\end{align}
Therefore,
\begin{equation}
\label{kads:int.R}
	\int\limits_0^\pi \rmd \theta \sin\theta \int\limits^R \rmd r\, \frac{Q}{\Delta_r} (F-m)
	= \ln R\, a^2 \sqrt{\Xi} A_{\text{log}} +\cdots~,
\end{equation}
where
\begin{equation}
\label{kads:a.log}
	A_{\text{log}} = \int\limits_0^\pi \rmd \theta\, \sin\theta \,\Delta_\theta^{-\frac32}
		\left(1-3\cos^2\theta +2 \frac{a^2}{L^2} \cos^4 \theta \right)~.
\end{equation}	   
After a change of the integration variable, this can be rewritten as 
\begin{equation}
\label{kadds:nodiv}
	A_{\text{log}} = \int\limits_{-1}^1 \rmd x
	\frac{1-3x^2 + 2\frac{a^2}{L^2} x^4}{\left(1-\frac{a^2}{L^2} x^2\right)^\frac32} 
	= \int\limits_{-1}^1 \rmd x\, \partial_x \frac{x-x^3}{\left(1-\frac{a^2}{L^2} x^2\right)^\frac12}=0~. 
\end{equation}
Therefore, the $\ln R$ term in \eqref{kads:int.R} vanishes, which proves that the complexity \eqref{kads:C} is finite in the limit  $R\to\infty$.

It is not possible to analytically compute the integrals in \eqref{kads:C} and to find a closed form for the complexity as a function of $\tau$. However, noting that \eqref{kads:C} depends on $\tau$ only through the integration extrema $r_m^\pm$ via \eqref{kads:rmeet}, the complexity growth rate can be obtained as
\begin{equation}
	\label{kads:Cdot}
	\frac{\rmd{\mathcal{C}}}{\rmd\tau} =\frac{1}{2\pi\hbar G\Xi} \int\limits_{0}^{\pi}\rmd\theta\sin\theta\left[F(r_m^-,\theta)-F(r_m^+,\theta)\right]~.
\end{equation}
Note that $r_m^\pm$ depend on $\theta$ (and on $\tau$). We can easily extract the following information from \eqref{kads:Cdot}. First the complexification rate is initially zero, because $r_m^+=r_m^-$ by symmetry for $\tau=0$. Second, it approaches a constant value at late times. To find this value, use $r_m^\pm\rightarrow r_\mp$ for $\tau\to\infty$ and the fact that $\Delta_r(r_\pm) = 0$. This implies 
\begin{align}
	\notag
	F(r_\pm, \theta) &=  m -\frac{3a^2 r_\pm \cos^2 \theta}{L^2} -\frac{a^2}{r_\pm} \\
	&= \frac{a^2 r_\pm}{L^2} \left( 1-3 \cos^2\theta\right) + \frac{r_\pm^3}{L^2} + r_\pm -m~.
\end{align}
Substituting in \eqref{kads:Cdot}, the late time complexity growth rate is obtained as 
\begin{equation}
	\label{kads:Cdot.limit}
	\lim_{\tau\rightarrow\infty}\frac{\rmd\mathcal{C}}{\rmd\tau} = 
	\frac{r_+^3-r_-^3+L^2(r_+-r_-)}{\pi\hbar G (L^2-a^2)}~.
\end{equation}
The result \eqref{kads:Cdot.limit} agrees with the result found in \cite{Bernamonti:2021jyu}.\footnote{The extra factor of two with respect to \cite{Bernamonti:2021jyu} is due to our definition of $\tau$.} Using \eqref{kads:MJ} and \eqref{kads:thermo}, it can be rewritten as \cite{Bernamonti:2021jyu}
\begin{equation}
	\label{kads:CdotlimitLloyd}
	\lim_{\tau\rightarrow\infty}\frac{\rmd\mathcal{C}}{\rmd\tau} = \frac{2}{\pi\hbar}\left[\left(M-\Omega_+J\right) - \left(M-\Omega_-J\right)\right]~,
\end{equation}
where $\Omega_+=\Omega$, and $\Omega_-$ is the analogous quantity associated with the inner horizon. This saturates the bound of \cite{Cai:2016xho}. 

We would like to establish, whether the \eqref{kads:Cdot.limit} is approached from below or above, so we need to calculate another derivative with respect to $\tau$. To make progress, first note that \eqref{kads:rmeet} implies
\begin{equation}
\label{kads:rpm.theta}
	\left. \frac{Q}{\Delta_r}\right|_{r_m^\pm} \frac{\rmd r_m^\pm}{\rmd\theta} + \left. \frac{P}{\Delta_\theta} \right|_{r_m^\pm} = \left. \frac{P}{\Delta_\theta} \right|_{R} \underset{R\to \infty}{=} 0~. 
\end{equation}
This can be used to substitute, in \eqref{kads:Cdot}, the fraction $\frac{P}{Q}$ that appears in $F(r_m^\pm,\theta)$ [c.f.\ \eqref{kads:F}]. Hence,
\begin{align}
\label{kads:F.pm}
	F(r_m^\pm,\theta)  &= m -\frac{3a^2r\cos^2\theta}{L^2} -\frac{a^2}{r} 
	-\frac{a^2}{L^2} \left(\frac{\rmd r}{\rmd\theta} \right) \sin\theta\cos\theta \\
\notag
	&\quad 
	-\frac12 \left[\Delta_r \partial_r -\Delta_\theta \left(\frac{\rmd r}{\rmd\theta}\right) \partial_\theta\right] \ln \left|\frac{\Pi}{r^2\Sigma^2}\right|~,
\end{align}
where $r=r_m^\pm$ is implied on the right hand side. Substituting \eqref{kads:F.pm} into \eqref{kads:Cdot} and integrating by parts the fourth term yields
\begin{equation}
	\label{kads:Cdot.new}
	\frac{\rmd{\mathcal{C}}}{\rmd\tau} =\frac{1}{2\pi\hbar G\Xi} \int\limits_{0}^{\pi}\rmd\theta\sin\theta\left[\tilde{F}(r_m^-,\theta)-\tilde{F}(r_m^+,\theta)\right]~,
\end{equation}
where 
\begin{align}
\notag
	\tilde{F}(r_m^\pm,\theta) &= m -\frac{a^2(r^2 + L^2)}{L^2r}
	-\frac12 \left[\Delta_r \partial_r -\Delta_\theta \left(\frac{\rmd r}{\rmd\theta}\right) \partial_\theta\right] \ln \left|\frac{\Pi}{r^2\Sigma^2}\right|\\
\label{kads:F.new}
	&= -m +r\left(1+\frac{r^2}{L^2}\right) 
	-\frac12 \left[\Delta_r \partial_r -\Delta_\theta \left(\frac{\rmd r}{\rmd\theta}\right) \partial_\theta\right] \ln \left|\frac{\Pi}{\Sigma^2}\right|~.
\end{align}
Again, $r=r_m^\pm$ is implied on the right hand side. 

We are now ready to take another derivative with respect to $\tau$, using again \eqref{kads:rmeet}. From \eqref{kads:F.new}, there will be terms involving 
\begin{equation}
\label{kads:mixed.deriv}
	\frac{\rmd}{\rmd \tau}\frac{\rmd}{\rmd \theta} r_m^\pm = \frac{\rmd}{\rmd \theta}\left. \left(\pm \frac{\Delta_r}{Q}\right)\right|_{r_m^\pm}~.
\end{equation}
Such terms can be integrated by parts. After a few lines, one finds
\begin{equation}
	\label{kads:Cddot}
	\frac{\rmd^2{\mathcal{C}}}{\rmd\tau^2} = \frac{1}{2\pi\hbar G\Xi} \int\limits_{0}^{\pi}\rmd\theta\sin\theta\left[G(r_m^-,\theta)+G(r_m^+,\theta)\right]~,
\end{equation}
where we have abbreviated
\begin{align}
\label{kads:Gdef}
	G(r,\theta) = -\frac{\Delta_r}{2Q} & \left[  2+\frac{6r^2}{L^2} - (\partial_r \Delta_r) \partial_r \ln \frac{\Pi}{\Sigma^2} 
	- \Delta_r \partial_r^2 \ln \frac{\Pi}{\Sigma^2} \right.\\
\notag
&\; \left.
		- (\partial_\theta \Delta_\theta) \partial_\theta \ln \frac{\Pi}{\Sigma^2} - \Delta_\theta \partial_\theta^2 \ln \frac{\Pi}{\Sigma^2} -\Delta_\theta (\partial_\theta\ln \sin\theta) \partial_\theta \ln \frac{\Pi}{\Sigma^2}\right]~.
\end{align}
Notice that the terms with mixed derivatives, $\partial_r\partial_\theta \ln\frac{\Pi}{\Sigma^2}$, have cancelled.

\begin{figure}[t]
	\begin{center}
		\includegraphics[width=0.5\textwidth]{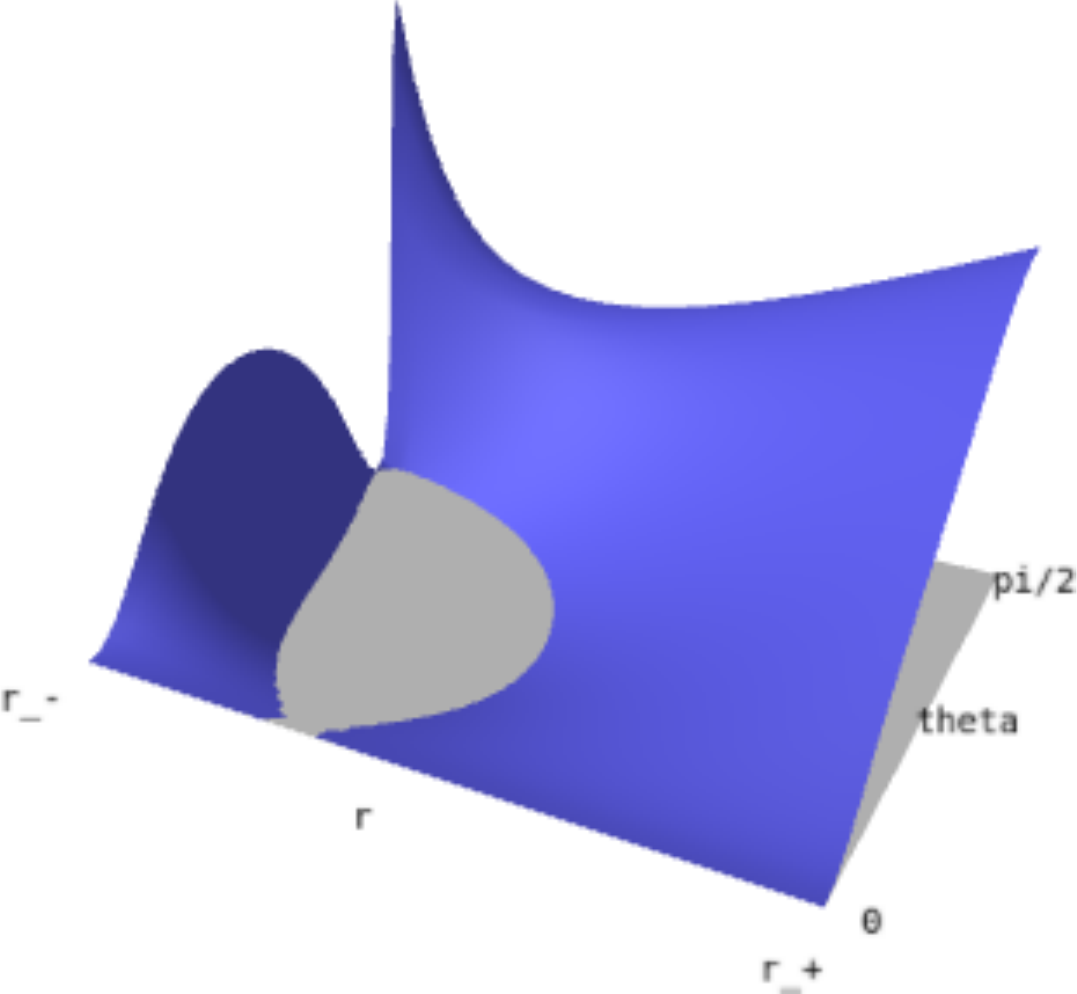}
	\end{center}
	\caption{Illustration of the term in the brackets in \eqref{kads:Gdef} for $r\in (r_-,r_+)$ and $\theta\in (0,\pi/2)$ with the parameters 
	$r_-=0.2$, $r_+=1.2$, $L=1$ ($a=0.9268$, $m=2.3373$). The gray plane corresponds to zero. The term in question is negative, where the gray plane is visible. \label{kads:3d-plot}}
\end{figure}

If the term in the brackets in \eqref{kads:Gdef} were non-negative for all $r\in (r_-,r_+)$ and $\theta\in (0,\pi)$, then it would follow immediately that \eqref{kads:Cddot} is positive. Unfortunately, this is not the case, as we illustrate in figure~\ref{kads:3d-plot}. It is impossible to make a general statement, because generically positive values at $r_m^-$ may be offset by possibly negative values at $r_m^+$,  and both, $r_m^-$ and $r_m^+$, are functions of $\theta$. One can show, however, that the term in question is non-negative in the late time limit, when $r_m^\mp\to r_\pm$.  
In this limit, the term in the brackets reduces to\footnote{We have eliminated $m$ in favour of $r_\pm$.} 
\begin{equation}
	\frac{a^2 (r_\pm^2+L^2)^2 [3r_\pm^4 + r_\pm^2a^2(1+\cos^2\theta) -a^4\cos^2\theta]^2 \sin^2\theta}{
	L^4 \Delta_\theta r_\pm^2 (r_\pm^2+a^2)^2 (r_\pm^2+a^2\cos^2\theta)^2}~,
\end{equation}
which is manifestly non-negative. This concludes the proof that the complexity growth rate approaches the limit \eqref{kads:CdotlimitLloyd} from below. 

\section{AdS-Vaidya spacetime}
\label{vda}

\subsection{Setup}

We now consider the  AdS-Vaidya spacetime, which gives a description of a process in which a shock wake collapses within an initial AdS geometry, and a one-sided AdS-Schwarzschild black hole is formed. The collapse can be caused, for example, by the insertion of a homogeneous shell of null fluid. The action functional for such a fluid coupled to gravity has extensively been discussed in \cite{Chapman:2018dem, Chapman:2018lsv} and references therein. The action is 
\begin{equation}
    \label{vda:action}
    I = \int\rmd x^{n+2}\sqrt{-g}\left(R - 2\Lambda + \mathcal{L}_{\text{fluid}}\right),
\end{equation}
where
\begin{equation}
    \label{vda:lagrangian}
    \mathcal{L}_{\text{fluid}} = \lambda g_{\mu\nu}l^\mu l^\nu + s l^\mu\partial_\mu\phi~.
\end{equation}
Here, $l^\mu$ is the velocity of the  null fluid, $g_{\mu\nu}$ is the metric, and the rest of the fields are auxiliary. In particular, $\lambda$ is a Lagrange multiplier that enforces the constraint $l^\mu l_\mu = 0$, $s$ can be interpreted as an entropy density, and $\phi$ plays the role of a velocity potential. The on-shell stress-energy tensor is
\begin{equation}
    \label{vda:stress}
    T_{\mu\nu} = 2\lambda l_\mu l_\nu~.
\end{equation}
By comparison with the stress tensor of a regular fluid we can identify the energy density of the null fluid as  $\epsilon =2 \lambda$.

The Vaidya metric in ingoing Eddington-Finkelstein coordinates is 
\begin{equation}
    \label{vda:metric}
    \rmd s^2 = -F(v,r)\rmd v^2 + 2\rmd v\rmd r + r^n\Omega_n^2~,
\end{equation}
with
\begin{equation}
    \label{vda:warp}
    F(v,r) = 1 + \frac{r^2}{L^2} - \frac{f_P(v)}{r^{n-1}}~.
\end{equation}
$f_P(v)$ is known as the profile function. From Einstein's equation we find
\begin{equation}
    \label{vda:energy}
    \lambda = \frac{n}{32\pi G}\frac{f'_P(v)}{r^n}~. 
\end{equation}
This shows that the derivative of the profile function is proportional to the energy density and, as such, should always be non-negative. When $f_P$ is a constant, there is no collapsing null fluid, and we simply have the AdS-Schwarzschild black hole geometry, or pure AdS if $f_P = 0$. Using the equation of motion one can also show that the null fluid Lagrangian \eqref{vda:lagrangian} vanishes on shell. Therefore, the null-fluid contributes only implicitly through the gravity action.

\subsection{Action in a causal diamond}

We focus on a causal diamond entirely contained in the null fluid region. Like we did in the previous sections, we will compute the action in the diamond, and then fix a parameterization. 

The action contribution of the four null boundaries is
\begin{align}
\label{vda:I.N}
	\frac{I_{N}}{2\Omega_n} &= 
	-n\int\limits_{r_1}^{r_2}\rmd r\,r^{n-1}\sigma_1
	-n\int\limits_{r_3}^{r_2}\rmd r\,r^{n-1}\sigma_2 
	-n\int\limits_{r_3}^{r_4}\rmd r\,r^{n-1}\sigma_3 
	-n\int\limits_{r_1}^{r_4}\rmd r\,r^{n-1}\sigma_4 \\
\notag
	&\quad 
	- r_1^n\left[\sigma_1(r_1)+\sigma_4(r_1)\right] + r_2^n\left[\sigma_1(r_2)+\sigma_2(r_2)\right] 
	- r_3^n\left[\sigma_2(r_3)+\sigma_3(r_3)\right] + r_4^n\left[\sigma_3(r_4)+\sigma_4(r_4)\right]~.
\end{align}
The bulk contribution is
\begin{equation}
    \label{vda:I.B}
    I_{B} = -\frac{2\Omega_n}{L^2}\int\limits_{v_3}^{v_1}\left(\rho_2^{n+1}-\rho_4^{n+1}\right)~.
\end{equation}
The functions $\rho(v)$ correspond to the radial coordinates of the null boundaries as a function of $v$ and satisfy $\frac{\rmd\rho}{\rmd v} = \frac12 F(v, \rho)$.
As before, we rewrite the bulk contribution using the identity
\begin{equation}
    \label{vda:id}
    \frac{\rho(v)}{L^2} = \frac{\rmd \ln |F(v, \rho(v))|}{\rmd v} - \frac{(n-1)f_P(v)}{2\rho^n} + \frac{f'_P(v)}{F(v,\rho(v))\rho^{n-1}}~,
\end{equation}
where the prime denotes a derivative with respect to $v$. After some manipulation, we find
\begin{align}
\label{vda:I.B2}
	\frac{I_B}{2\Omega_n} 
		&= n a_u\int\limits_{r_1}^{r_2}\rmd r\,r^{n-1}\ln |F(v(r), r)| 
		 + n a_u\int\limits_{r_3}^{r_4}\rmd r\,r^{n-1}\ln |F(v(r), r)| \\
	\notag &\quad
		+ n a_v\int\limits_{r_3}^{r_2}\rmd r\,r^{n-1}\ln |F(v(r), r)| 
		+ n a_v\int\limits_{r_1}^{r_4}\rmd r\,r^{n-1}\ln |F(v(r), r)| \\
	\notag &\quad 
	-\int\limits_{v_3}^{v_1}\rmd v\,f_P'(v)\left(\frac{\rho_2}{F(v,\rho_2)}- \frac{\rho_4}{F(v,\rho_4)}\right)\\
	\notag &\quad 
		+ r_1^n\ln |F(v_1,r_1)|-r_2^n\ln |F(v_1,r_2)| + r_3^n\ln |F(v_3,r_3)|-r_4^n\ln |F(v_3,r_4)|~,
\end{align}
with constants $a_u$ and $a_v=1-a_u$. 

We do not write the contribution of the corners since, just like in the AdS-Schwarzschild and the RN-AdS cases, these are cancelled by the surface terms in \eqref{vda:I.B2} and \eqref{vda:I.N}. From \eqref{vda:metric} and \eqref{vda:warp} one can see that the function $F(r,v)$ plays the role of the blackening factor for the Vaidya spacetime. Therefore, we fix
\begin{equation}
\label{vda:choice}
\begin{aligned}
\text{on $N_1$:}\quad \sigma_1(\lambda) &= a_u \ln |F(v(\lambda),r(\lambda))|~, \qquad 
\text{on $N_3$:}\quad \sigma_3(\lambda)  = a_u \ln |F(v(\lambda),r(\lambda))|~,\\
\text{on $N_2$:}\quad \sigma_2(\lambda) &= a_v \ln |F(v(\lambda),r(\lambda))|~, \qquad 
\text{on $N_4$:}\quad \sigma_4(\lambda)  = a_v \ln |F(v(\lambda),r(\lambda))|~.
\end{aligned}
\end{equation}
The total action on the causal diamond is then
\begin{align}
    \label{vda:CD}
    \frac{I}{2\Omega_n} &=  
	\int\limits_{v_1}^{v_3}\rmd v\,f_P'(v)\left[\frac{\rho_2}{F(v,\rho_2)}- \frac{\rho_4}{F(v,\rho_4)}\right]~.
\end{align}
One can check that the parameterization \eqref{vda:choice} is compatible with the criterion of vanishing action in a static vacuum. Again, although we have not written it explicitly, the corner contributions vanish separately. 

\subsection{Complexity=Action}

\begin{figure}[t]
	\begin{center}
		\includegraphics[width=.40\textwidth,align=c]{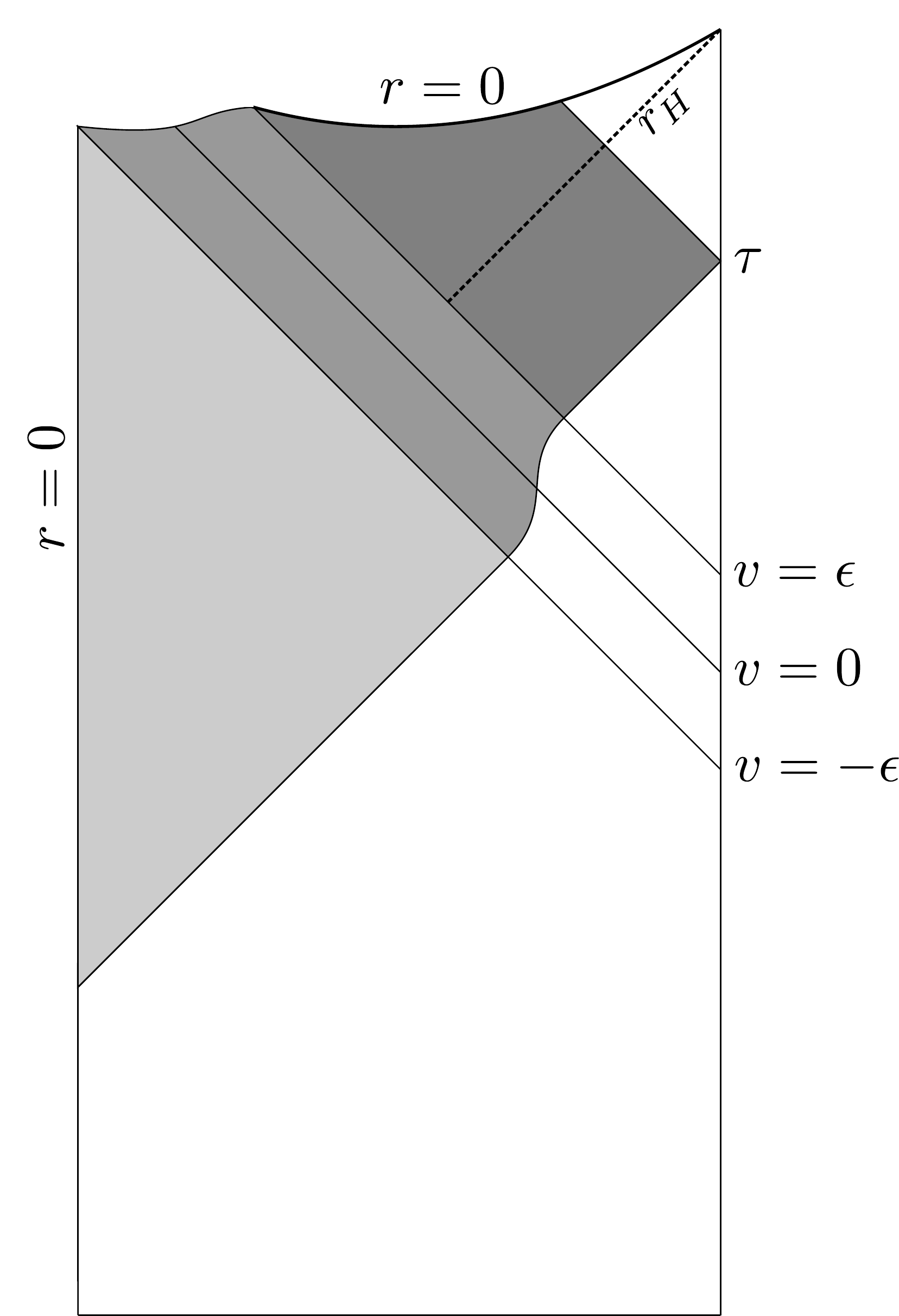}%
		\caption{WdW patch for Vaidya spacetime. The shell of null fluid is centred on $v=0$ and has thickness $2\delta$.\label{fig:vda.WdW}}
	\end{center}
\end{figure}

We suppose for simplicity that the collapsing shell of null fluid is centered at $v = 0$ and has thickness $2\delta$, as illustrated in Fig.~\ref{fig:vda.WdW}. 
We want to model a situation in which, after the collapse of a shell of null fluid in AdS space, a one-sided Schwarzschild black hole is formed. The profile function of the fluid must then interpolate between the two regimes, \ie $f_P(-\delta) = 0$ and $f_P(+\delta) = \omega^{n-1}$, with the condition that $f_P'>0$.
We can divide the WdW patch in three sub-regions. As shown in \cite{Mounim:2021bba}, using the appropriate parameterization, the sub-region with pure AdS geometry has zero action, while the black hole region contributes\footnote{The result appears to be one half of what was found in \cite{Mounim:2021bba} for an eternal AdS-Schwarzschild black hole, because the black hole formed by a collapse is one-sided.}
\begin{equation}
    \label{vda:I.WdW.BH}
    I^{\text{BH}} = 2n\Omega_n\omega^{n-1}(\tau - \delta)~.
\end{equation}

To evaluate the action on the sub-region of the WdW patch filled by the null fluid we, use the parameterization \eqref{vda:choice} with $a_v = 1$. We remark that the functions $\sigma$ on the null boundaries that traverse different sub-regions of the WdW patch, should be continuous.
The contribution of the null boundary is then 
\begin{equation}
    \label{vda:I.WdW.N}
    I_{N} = 2\Omega_n\int\limits_{-\delta}^{+\delta}\rmd v\left[\frac{\rho_s^{n+1}}{L^2} + \frac{(n-1)f_P(v)}{2} - f'_P\frac{\rho_s}{F(v,\rho_s)}\right]~.
\end{equation}
$\rho_s(v)$ is radial coordinate of the boundary along the shell of null fluid. 
The bulk contribution of this sub-region is
\begin{equation}
    \label{vda:I.WdW.B}
    I_{B} = -\frac{2\Omega_n}{L^2}\int\limits_{-\delta}^{+\delta}\rmd v\,\rho_s^{n+1}~.
\end{equation}
Finally, the contribution of the space-like boundary at $r=0$ is
\begin{equation}
    \label{vda:I.WdW.S}
    I_{S} = (n+1)\Omega_n\int\limits_{-\delta}^{+\delta}\rmd v\,f_P(v)~. 
\end{equation}
Adding up the contributions \eqref{vda:I.WdW.N}, \eqref{vda:I.WdW.B} and \eqref{vda:I.WdW.S}, we find the total action of the fluid sub-region of the WdW,
\begin{equation}
    \label{vda:I.WdW.fluid}
    \frac{I^{\text{fluid}}}{2\Omega_n} = \int\limits_{-\delta}^{+\delta}\rmd v\left(nf_P(v) - f'_P\frac{\rho_s}{F(v,\rho_s)}\right)~.
\end{equation}
Now, we can write the action of the whole WdW patch
\begin{equation}
    \label{vda:I.WdW}
    I = 
    \begin{cases}
    	0 \qquad &\text{for $\tau<-\delta$,}\\
    	2\Omega_n\int\limits_{-\delta}^{\tau}\rmd v\left(nf_P(v) - f'_P\frac{\rho_s}{F(v,\rho_s)}\right) &\text{for $|\tau|\leq\delta$,}\\
    	2\Omega_n\left[n\omega^{n-1}(\tau - \delta) + \int\limits_{-\delta}^{+\delta}\rmd v\left(nf_P(v) - f'_P\frac{\rho_s}{F(v,\rho_s)}\right)\right] &\text{for $\tau>\delta$~.}
    \end{cases}
\end{equation}

The action \eqref{vda:I.WdW} depends on the shape of the profile function. As a particular case, we study the infinitely thin shell limit, \ie $\delta\to 0$. In this limit, the profile function can be approximated by $f_P(v) = \frac12\omega^{n-1}( 1 + \frac{v}{\delta})$. Furthermore, since the radial coordinate is continuous across the shell, we can write $\rho_s(v) = r_s + O(\delta)$, where $r_s = \rho_s(0)$. Then, the integral in the third case of \eqref{vda:I.WdW} becomes
\begin{equation}
    \int\limits_{-\delta}^{+\delta}\rmd v\left(nf_P(v) - f'_P\frac{\rho_s}{F(v,\rho_s)}\right) = -\int\limits_{-\delta}^{+\delta}\rmd v\frac{r_s^{n}}{v} + O(\delta) = O(\delta)~.
\end{equation}
This shows that in the infinitely thin shell limit, the complexity of Vaidya spacetime is simply
\begin{equation}
    \label{vda:I.WdW.thin}
    \mathcal{C} = 
    \begin{cases}
    0 \qquad &\text{for $\tau<0$,}\\
    \frac{2M}{\pi\hbar}\tau &\text{for $\tau\geq 0$~.}
    \end{cases}
\end{equation}
As we could have expected, the complexity is zero before the collapse, and afterwards we have the linearly growing black hole contribution.
\section{Conclusions}
\label{conc}

In this paper, we have reconsidered the computation of holographic complexity in the $C=A$ approach for different types of asymptotically AdS black holes. The difference with previous works is that we do not require the gravitational action to be invariant under reparameterization of the null boundaries, and we do not add the covariance counter term. On the bulk side, the parameterization of the null boundaries describes the heat content on these boundaries \cite{Chakraborty:2019doh}. On the CFT side, we interpret the freedom in the choice of the parameterization as the choice of the details involved in the definition of computational complexity, such as the reference state and the set of elementary gates. Following our earlier proposal \cite{Mounim:2021bba}, we choose the parameterization according to the principle that \emph{the action in any vacuum stationary causal diamond vanishes}. Incidentally, this criterion seems to imply that \emph{the corner contributions to the action of a causal diamond vanish separately}. This is actually easier to implement than the original criterion.

In the considered cases, namely RN-AdS, rotating BTZ black holes and shock wave geometries, we find that the complexity growth rate at late times agrees with previous results in the literature. There are, however, important differences between our and the earlier results. First, within our approach, the complexity turns out to be finite when the cut-off $R$ is removed. This suggests that the action of the WdW patch, calculated with a parameterization such that the corner terms vanish, may not be dual to computational complexity, but to a sort of operator complexity \cite{Parker:2018yvk}. Second, we were able to prove, except for the Kerr-AdS case, that the complexity growth rate always satisfies a generalized bound and saturates it at late times. In contrast, with the standard $C=A$ procedure using an affine parameterization of the null boundaries and adding the counter term, this bound is typically violated. 

Therefore, we have confirmed that the application of our criterion successfully computes the complexity of a large class of black holes in Einstein-Hilbert gravity and provides an important improvement over the standard $C=A$ procedure with the counter term. It would be interesting to investigate the application to more general gravitational theories, such as Gauss-Bonnet gravity or higher derivative theories.

\section*{Acknowledgements}
This work was supported partly by the INFN, research initiative STEFI.

\bibliographystyle{JHEPnotes}
\bibliography{adscft,complexity,nsf}

\end{document}